\documentclass[12pt]{article}
\usepackage{latexsym}
\usepackage{amsmath}
\usepackage{amssymb}
\usepackage{braket}
\catcode`\@=11
\newcount\hour
\newcount\minute
\newtoks\amorpm \hour=\time\divide\hour by 60\minute
=\time{\multiply\hour by 60 \global\advance\minute by-\hour}
\edef\standardtime{{\ifnum\hour<12 \global\amorpm={am}%
        \else\global\amorpm={pm}\advance\hour by-12 \fi
        \ifnum\hour=0 \hour=12 \fi
        \number\hour:\ifnum\minute<10
        0\fi\number\minute\the\amorpm}}
\edef\militarytime{\number\hour:\ifnum\minute<10
0\fi\number\minute}
\def\draftlabel#1{{\@bsphack\if@filesw {\let\thepage\relax
   \xdef\@gtempa{\write\@auxout{\string
      \newlabel{#1}{{\@currentlabel}{\thepage}}}}}\@gtempa
   \if@nobreak \ifvmode\nobreak\fi\fi\fi\@esphack}
        \gdef\@eqnlabel{#1}}
\def\@eqnlabel{}
\def\@vacuum{}
\def\marginnote#1{}
\def\draftmarginnote#1{\marginpar{\raggedright\scriptsize\tt#1}}
\def\draft{
        \pagestyle{plain}
        \overfullrule=2pt
        \oddsidemargin -.1truein
        \def\@oddhead{\sl \phantom{\today\quad\militarytime} \hfil
        \smash{\Large\sl DRAFT} \hfil \today\quad\militarytime}
        \let\@evenhead\@oddhead
        \let\label=\draftlabel
        \let\marginnote=\draftmarginnote
        \def\ps@empty{\let\@mkboth\@gobbletwo
        \def\@oddfoot{\hfil \smash{\Large\sl DRAFT} \hfil}
        \let\@evenfoot\@oddhead}
        \def\@eqnnum{(\theequation)\rlap{\kern\marginparsep\tt\@eqnlabel}%
        \global\let\@eqnlabel\@vacuum}  }

\renewcommand{\theequation}{\thesection.\arabic{equation}}
\renewcommand{\thefootnote}{\fnsymbol{footnote}}

\def\appendix#1{\addtocounter{section}{1}\setcounter{equation}{0}
\renewcommand{\thesection}{\Alph{section}}
\section*{Appendix \thesection\protect\indent \parbox[t]{11.15cm}{#1}}
\addcontentsline{toc}{section}{Appendix \thesection\ \ \ #1}}

\def \bi{\bibitem}

\def\be{\begin{equation}}
\def\ee{\end{equation}}

\catcode`\@=11

\def\bea{\begin{eqnarray}}
\def\eea{\end{eqnarray}}
\def\beann{\begin{eqnarray*}}
\def\eeann{\end{eqnarray*}}
\def\beq{\begin{equation}}
\def\eeq{\end{equation}}
\def\ba{\begin{array}}
\def\ea{\end{array}}
\def\ben{\begin{enumerate}}
\def\een{\end{enumerate}}

 \def\be{\begin{equation}}
\def\ee{\end{equation}}

\def \ba {{\bar{\alpha}}}

\def\be{\begin{equation}}
\def\ee{\end{equation}}

\def \bi {\bibitem}

\begin{document}
\date{November 2002}
\begin{titlepage}
\begin{center}
\vspace*{-1.0cm}

\vspace{2.0cm} {\Large \bf M-theory  backgrounds with 30 Killing spinors are maximally supersymmetric} \\[.2cm]

\vspace{1.5cm}
 {\large  U. Gran$^1$, J. Gutowski$^2$ and  G. Papadopoulos$^2$}

\vspace{0.5cm}

${}^1$ Fundamental Physics\\
Chalmers University of Technology\\
SE-412 96 G\"oteborg, Sweden\\

\vspace{0.5cm}
${}^2$ Department of Mathematics\\
King's College London\\
Strand\\
London WC2R 2LS, UK\\

\vspace{0.5cm}

\end{center}

\vskip 1.5 cm
\begin{abstract}
We show that all M-theory backgrounds which admit more than 29 Killing spinors are maximally supersymmetric.
In particular, we find that the
supercovariant curvature of all  backgrounds  which preserve 30 supersymmetries, subject to field equations and Bianchi identities,
vanishes, and that there are no such solutions which arise as
discrete quotients of maximally supersymmetric backgrounds.

\end{abstract}

\end{titlepage}
\newpage
\setcounter{page}{1}
\renewcommand{\thefootnote}{\arabic{footnote}}
\setcounter{footnote}{0}

\setcounter{section}{0}
\setcounter{subsection}{0}

\section{Introduction}

In recent years  much progress has been made towards understanding  supersymmetric  M-theory backgrounds.
In particular, the maximally supersymmetric backgrounds have been classified in \cite{georgejose}, and the Killing spinor
equations for one Killing spinor have been solved in \cite{pakis}. More rapid development  took place with the introduction
of the spinorial geometry technique \cite{mtheor1} for solving the Killing spinor equations. This allowed the solution of the Killing spinor
equations for  more than one Killing spinor \cite{mtheor1} and initiated the exploration of type II
backgrounds with near maximal number of supersymmetries \cite{n31iib, d11preon}. In particular, it has been shown that
IIB backgrounds which admit more than 28 Killing spinors are maximally supersymmetric \cite{n31iib, more28iib}, and that the plane wave solution of \cite{bena}
 is the unique \cite{28iib}
local geometry which admits 28 supersymmetries. Moreover, it has been demonstrated that M-theory backgrounds which admit
31 Killing spinors are maximally supersymmetric \cite{d11preon, josegadhia}.   A similar result holds for type IIA backgrounds \cite{bandos}
which was proven using a different technique.

The above results on nearly maximally supersymmetric backgrounds in M-theory and IIB supergravity have illuminated
some long standing questions regarding the structure of supersymmetric backgrounds in theories with 32 supercharges.
In particular, the results obtained are in agreement with a conjecture in \cite{duff} about the
number of supersymmetries preserved by M-theory and type II supergravity backgrounds. They are also consistent
with the homogeneity conjecture of \cite{figueroa} which postulates that all solutions of supergravity theories which
preserve more than $1/2$ of supersymmetry are homogeneous.

 Another question that the results on nearly maximal
supersymmetric backgrounds  elucidate is whether there are gravitational backgrounds for every BPS
state of  the  supersymmetry algebra with  brane charges \cite{townsend}. To explain this, it is expected that for every BPS state
there is a supergravity background with the same asymptotic charges as those that characterize the state. This is because
 such states are massive and so self-gravitate. BPS states of supersymmetry algebras with brane charges can be found that
 preserve nearly maximal numbers of supersymmetries \cite{bandosalg}. In particular, those  which preserve 31 supersymmetries have been called preons.
 However, as we have mentioned
there are no solutions of supergravity theories with this number of supersymmetries. The reason behind this is that
in the supergravity calculation, apart from the kinematical effects which are represented to some extent by the Killing spinor equations,
the dynamics is also important. In particular the field equations and the Bianchi identities are used to
rule out the existence of such backgrounds\footnote{If only the
restrictions on the fields imposed  by the Killing spinor equations are taken into account, then there may be configurations
which preserve 31 supersymmetries. This is because the   holonomy of the supercovariant connection of 11-dimensional and IIB supergravities
is in $SL(32,{\mathbb{R}})$
\cite{hull, duff2, tsimpisgp}  but such configurations do not satisfy the field equations and Bianchi identities.}.
Moreover whenever nearly maximally supersymmetric backgrounds are known to exist, they
 are typically plane waves and do not admit appropriate asymptotic brane
charges in order to be identified with the BPS states which preserve the same number of supersymmetries.
This incompatibility between the supersymmetry algebra considerations and supergravity calculations is not fully understood and affects
many BPS states which preserve more than $1/2$ of the supersymmetry.

 In this paper, we shall extend the results on the existence of nearly maximally supersymmetric solutions
 of M-theory by showing that all solutions with 30 Killing spinors are maximally supersymmeric.
The proof relies
on the use of the gauge symmetry of 11-dimensional supergravity to choose the two normals to the 30-dimensional plane of Killing spinors.
This treatment is similar to that which has been used to examine other nearly maximally supersymmetric solutions in \cite{n31iib, d11preon}.
Putting the two normals in a canonical form and using the orthogonality condition of
the $Spin(10,1)$ invariant metric on the space of spinors, we  choose the
30 Killing spinors. Then the integrability condition of the Killing spinor equations, which involves the supercovariant
curvature, is solved. It is shown that subject to field equations and Bianchi identities, all components
of the supercovariant curvature vanish. This establishes that all backgrounds with 30 supersymmetries are locally isometric
to the maximally supersymmetric solutions of 11-dimensional supergravity. To complete the proof, it remains
to show that there are no discrete quotients of maximally supersymmetric backgrounds which preserve 30 supersymmetries.
This is also established using the general method proposed in \cite{josesimon} and applied in \cite{josegadhia} to show a
similar result for the case of 31 supersymmetries.

We also investigate the existence of plane wave solutions in M-theory which preserve 28 supersymmetries.
This is motivated by the result in IIB supergravity, mentioned above,
that this solution is unique  and not locally  maximally supersymmetric.
Moreover it preserves the highest fraction of supersymmetry other than maximal.
The possibility of the existence of such solutions in M-theory has been raised in \cite{gunaydin} with the construction
of  a plane wave superalgebra  with 28 odd generators and even subalgebra  $(\mathfrak{so}(3)\oplus \mathfrak{su}(3)\oplus \mathfrak{u}(1))\oplus_s
\mathfrak{H}_9$,
 where $\mathfrak{H}_9$ is a Heisenberg
algebra and $\oplus_s$ denotes semi-direct sum.
We find that the plane wave solution which has as bosonic symmetry\footnote{The superalgebra   considered here is the symmetry algebra,
 which includes the isometries,  of the solution in the spirit of \cite{fofgp, maxsusy} and it should  not be confused with
asymptotic supersymmetry algebra with  brane charges
mentioned earlier.}
 the  subalgebra $(\mathfrak{so}(3)\oplus \mathfrak{su}(3)\oplus \mathfrak{u}(1))\oplus_s
\mathfrak{H}_9$ actually
preserves either 16,  20 or 32, but not 28, supersymmetries depending on the choice of parameters\footnote{
The apparent absence of a plane wave solution admitting a symmetry superalgebra with 28 odd generators,
as discussed in \cite{gunaydin}, is puzzling.}. The $N=20$ solution has been constructed before in \cite{hg}.

This paper is organized as follows. In section two, we state the identities on the components of the supercurvature
implied by the field equations and Bianchi identities of 11-dimensional supergravity, the ${\cal R}$-identities.
In section 3, we give the canonical forms
of the two normals to the Killing spinors. In section 4, we solve the ${\cal R}$-identities for backgrounds with 30 Killing spinors.
In section 5, we show that the supercurvature of backgrounds with 30 supersymmetries vanishes using in addition
the explicit dependence of the supercurvature on the fundamental fields. In section 6, we complete the proof by demonstrating that there
are no backgrounds with 30 supersymmetries which arise as discrete quotients of maximally supersymmetric backgrounds.  In section
7 we investigate a class of plane wave solutions conjectured to preserve 28 supersymmetries,
and in section 8 we give our conclusions.
In appendices  A, B and C, we present details of the computation for the choice
of normal spinors and for the analysis of ${\cal R}$-identities. In appendix D, we investigate the existence of plane wave
solutions with 28 supersymmetries.

\section{The Integrability Conditions}

The bosonic fields of 11-dimensional supergravity \cite{julia}  are a metric $g$ and a 4-form field strength $F$.
The first part the proof that all M-theory backgrounds with 30 supersymmetries are maximally supersymmetric relies
on the properties of the curvature of the supercovariant connection. In particular,  the integrability condition
of the Killing spinor equation, ${\cal D}\epsilon^r=0$  is
\begin{equation}
\label{intcon}
  {\cal R}_{MN} \epsilon^r=[{\cal D}_M, {\cal D}_N]\epsilon^r = \sum^5_{k=1} {1\over k!}(T_{MN}^k)_{A_1A_2\dots A_k}
(\Gamma^{A_1A_2\dots A_k}) \epsilon^r =0
\end{equation}
where $\{ \epsilon^r \}$ for $r=1, \dots , 30$ is a basis for the Killing
spinors, and ${\cal R}$ is the supercovariant curvature. The (real) components  $T$ of ${\cal R}$ depend on the physical fields and their
derivatives, and some of them contain the Riemann curvature of spacetime. Their precise expressions
are given in \cite{georgejose}.

An essential part of the proof is to show that if there are 30 linearly independent Killing spinors, then ${\cal R}=0$. This
will demonstrate that the backgrounds with 30 supersymmetries are locally maximally supersymmetric.
To show this, one has to implement the field equations and Bianchi identities of 11-dimensional supergravity as well as utilize
the explicit dependence of ${\cal R}$  on the physical fields. In turn, some of these conditions  can be expressed
as relations on the components $T$ of ${\cal R}$

\begin{eqnarray}
\label{tcon1}
    && (T^1_{MN})^N = 0~,~~~  (T^2_{MN})_P{}^N = 0~,~~~(T^1_{MP_1})_{P_2} + \frac{1}{2} (T^3_{MN})_{P_1 P_2}{}^N =  0~, \cr
    && (T^2_{M[P_1})_{P_2 P_3]} - \frac{1}{3} (T^4_{MN})_{P_1 P_2 P_3}{}^N =  0~,~~~
    (T^3_{M[P_1})_{P_2 P_3 P_4]} + \frac{1}{4} (T^5_{MN})_{P_1 \cdots P_4}{}^N =  0~, \cr
    && (T^4_{M[P_1})_{P_2 \cdots P_5]} - \frac{1}{5 \cdot 5!} \epsilon_{P_1 \cdots P_5}{}^{Q_1 \cdots Q_6}
      (T^5_{M Q_1})_{Q_2 \cdots Q_6}  =  0~
 \end{eqnarray}
 \begin{eqnarray}
 \label{tcon2}
  && (T^1_{MN})_P = (T^1_{[MN})_{P]} \,, \qquad
  (T^2_{MN})_{PQ} = (T^2_{PQ})_{MN} \,, \qquad
  (T^3_{[MN})_{PQR]} = 0 \ ,
 \end{eqnarray}

\begin{equation}
\label{tcon3}
(T^3_{(M_1 |(N_1})_{N_2)| M_2 )M_3} =0 \ ,
\end{equation}
and
\begin{equation}
\label{tcon3b}
 (T^4_{(M_1 |(N_1})_{N_2) |M_2 )M_3 M_4} =  0\ .
\end{equation}
For convenience, we shall refer to (\ref{tcon1})-(\ref{tcon3b}) as the {\it supercurvature identities} or ${\cal R}$-identities
for short.
In order to analyse the $N=30$ solutions, it is particularly useful to note the following conditions, which relate
the 4-form field strength to the $T^i$:

\begin{eqnarray}
\label{algcon1}
F_{[N_1 N_2 N_3 N_4} F_{N_5 N_6 N_7 N_8]} = {6 \over 35}
\epsilon_{N_1 N_2 N_3 N_4 N_5 N_6 N_7 N_8}{}^{M_1 M_2 M_3} (T^1_{M_1 M_2})_{M_3}
\end{eqnarray}
\begin{eqnarray}
\label{algcon2}
F_{M [Q_1 Q_2 Q_3} F_{|N| Q_4 Q_5 Q_6]}
&=& {1 \over 5!} \epsilon_{Q_1 Q_2 Q_3 Q_4 Q_5 Q_6}{}^{N_1 N_2 N_3 N_4 N_5}
\bigg( {9 \over 10} (T^5_{MN})_{N_1 N_2 N_3 N_4 N_5}
\nonumber \\
&+&{3 \over 2} (T^5_{[M| N_1})_{N_2 N_3 N_4 N_5 |N]} -2
(T^5_{[M| L})_{N_1 N_2 N_3 N_4}{}^L \eta_{N_5| N]}
\nonumber \\
&-&4 \eta_{M N_1} \eta_{N N_5} (T^5_{L_1 L_2})_{N_2 N_3 N_4}{}^{L_1 L_2} \bigg)
\end{eqnarray}
\begin{eqnarray}
\label{algcon3}
F_{L [N_1 N_2 N_3} F_{N_4] MN}{}^L
&=& {9 \over 2} (T^4_{MN})_{N_1 N_2 N_3 N_4}
+6 (T^4_{NL})_{[N_1 N_2 N_3}{}^L \eta_{N_4] M}
\nonumber \\
&-&6  (T^4_{ML})_{[N_1 N_2 N_3}{}^L \eta_{N_4] N}
\end{eqnarray}

We use the method introduced in  \cite{d11preon}
to solve the integrability conditions ({\ref{intcon}}).
 In particular, we introduce the normals $\nu^p$ to the Killing spinors with respect
to the Majorana inner product $B$ and write
\begin{equation}
\label{supercurv1}
{\cal R}_{MN,ab}= u_{MN,ip}\,\eta^i_a
\nu^p_b
\end{equation}
where $a, b$ are spinor indices\footnote{We follow the form and spinor conventions in  \cite{mtheor1, mtheor2}.}, and $\{ \eta^i \}$ for $i=1, \dots , 32$
is a canonical Majorana basis, either in the timelike or null basis,
as described in \cite{mtheor1, mtheor2}, and $u$'s are
real spacetime functions. Clearly ${\cal R}$ expressed as in (\ref{supercurv1}) satisfies the integrability condition (\ref{intcon}).
Next,
on using the spinor identity
\begin{eqnarray}
\eta_a \theta_b=  \frac{1}{32}\sum^5_{k=0} {(-1)^{k+1}\over k!} B(\eta,
\Gamma_{A_1A_2\dots A_k}\theta)\,  (\Gamma^{A_1A_2\dots
A_k})_{ab}~,
\end{eqnarray}
one finds that the components $T$ are expressed in terms of the $u$'s as
\begin{eqnarray}
(T_{MN}^k)_{A_1A_2\dots A_k}= \frac{(-1)^{k+1}}{32} u_{MN,ip}\,
B(\eta^i, \Gamma_{A_1A_2\dots A_k}\nu^p)
\label{supcurv2} \ .
\end{eqnarray}

Substituting these expressions for $T$'s back into the ${\cal R}$-identities, one obtains conditions on
 $u$'s. In particular, if the ${\cal R}$-identities imply that $u=0$, then $T=0$
and the associated solutions are locally maximally supersymmetric.

In addition to the conditions on the $u$'s imposed by the ${\cal R}$-identities, there is also the restriction
\begin{eqnarray}
\label{tcon4}
u_{MN,iq}\, B(\eta^i,\nu^q)=0~.  \label{supcon2}
\end{eqnarray}
This is because the (reduced) holonomy of the supercovariant connection is contained in $SL(32,{\mathbb{R}})$ \cite{hull, duff2} rather than
$GL(32,{\mathbb{R}})$. The above condition is the requirement that the trace of the supercovariant curvature vanishes.

\section{Normal Spinors}

Further progress to proving whether ${\cal R}=0$ for backgrounds with 30 supersymmetries depends on the use
of gauge symmetry $Spin(10,1)$ of 11-dimensional supergravity to choose the two normals $\nu^1$ and $\nu^2$ of the Killing spinors.
The first normal can be chosen as in \cite{d11preon}. In particular, there are two inequivalent orbits of $Spin(10,1)$
in the space of Majorana spinors with isotropy groups $SU(5)$ and $Spin(7) \ltimes {\mathbb{R}}^9$.  A representative of the $SU(5)$
orbit is
\begin{equation}
\nu^1 = 1+e_{12345}~,
\end{equation}
and a representative of the   $Spin(7) \ltimes {\mathbb{R}}^9$ orbit is
\begin{equation}
\nu^1 = 1+e_{1234} \ .
\end{equation}
It is essential to note that the representatives of the two different orbits have been expressed in {\it two   different} bases.
The representative of the $SU(5)$ orbit has been written in the time-like basis while the representative of the $Spin(7) \ltimes {\mathbb{R}}^9$
has been written in the null basis, for the definition of these spinor bases see \cite{mtheor1, mtheor2}. Note that the
1-form spinor bi-linear of the $SU(5)$ invariant normal is time-like while the same form of the $Spin(7) \ltimes {\mathbb{R}}^9$ invariant norma is null.
 In what follows,
we shall use the remaining gauge symmetry  to choose the second {\it linearly independent}
normal $\nu^2$ to the Killing spinors. We shall label the two cases with the isotropy groups of the first normal.

\subsection{$SU(5)$}

Suppose that the first normal is $\nu^1 = 1+e_{12345}$. To choose the second normal up to  $SU(5)$ transformations that leave invariant
$\nu^1$, we  first note that the most general form of $\nu^2$
in the time-like spinor basis of  \cite{mtheor1} is
\begin{eqnarray}
\nu^2 &=& \alpha .1 + {\bar{\alpha}} e_{12345} + \beta^k e_k +{1 \over 4!} (\star {\bar{\beta}})^{m_1 m_2 m_3 m_4} e_{m_1 m_2 m_3 m_4}
\nonumber \\
&+& {1 \over 2} \sigma^{n_1 n_2} e_{n_1 n_2} -{1 \over 3!} (\star {\bar{\sigma}})^{k_1 k_2 k_3} e_{k_1 k_2 k_3}~,
\end{eqnarray}
where here $k,m,n=1, ... ,5$ and $\alpha, \beta^k, \sigma^{mn}$ are in general complex valued, and $\star$ denotes the Hodge
dual on ${\mathbb{R}}^5$. Then we decompose the Majorana representation of $Spin(10,1)$ under $SU(5)$ and appropriately choose representatives
for the orbits of isotropy groups. The procedure has been explained in detail in appendix A.
It turns out that there are two cases to consider, according to whether  $\beta=0$ or $\beta \neq 0$.
In the  $\beta \neq 0$ case, the second normal spinor can be chosen as
\begin{eqnarray}
\nu^2 &=&  ix (1-e_{12345}) + e_1 +e_{2345}
+ \sigma^{12} (e_{12}-e_{345}) + \sigma^{34} (e_{34}-e_{125}) + \sigma^{45}(e_{45}-e_{123})
\nonumber \\
&+& \sigma^{23} e_{23} - {\bar{\sigma}}^{23} e_{145}~,
\label{snorm1}
\end{eqnarray}
where $x, \sigma^{12}, \sigma^{34}, \sigma^{45}$ are real spacetime functions.

In  $\beta=0$ case , the second normal spinor can be chosen as
\begin{equation}
\nu^2 = ix (1-e_{12345}) + \sigma^{12}(e_{12}-e_{345}) + \sigma^{34}(e_{34}-e_{125})~,
\label{snorm2}
\end{equation}
where $x, \sigma^{12}, \sigma^{34}$ are real spacetime functions.

\subsection{$Spin(7)\ltimes{\mathbb{R}}^9$}

Suppose that the first normal spinor is $\nu^1=1+e_{1234}$. To choose the second normal
up to $Spin(7) \ltimes {\mathbb{R}}^9$ that leave invariant $\nu^1$, we first note that the most general form of $\nu^2$
in the null basis of \cite{mtheor2} is

\begin{eqnarray}
\nu^2 &=& \alpha 1 + {\bar{\alpha}} e_{1234} + w e_5 + {\bar{w}} e_{12345}
+ \tau^j e_j -{1 \over 3!} (\star {\bar{\tau}})^{n_1 n_2 n_3} e_{n_1 n_2 n_3}
\cr
&+& \psi^j e_{j5} -{1 \over 3!} \star {\bar{\psi}}^{n_1 n_2 n_3} e_{n_1 n_2 n_3 5}
+{1 \over 2} (A^{ij} - \star {\bar{A}}^{ij}) e_{ij} +{1 \over 2} (B^{ij} - \star {\bar{B}}^{ij}) e_{ij5}~,
\end{eqnarray}
where here $i, j, n=1, ..., 4$, $\alpha, w, \tau^i, \psi^i, A^{ij}, B^{ij}$ are complex valued
and $\star$ denotes the Hodge dual on ${\mathbb{R}}^4$.

After a detailed analysis which can be found in appendix A, the second normal can be written in
one of four possible canonical forms:

\begin{equation}
\label{nullx1}
\nu^2 =c_1 (e_5 + e_{12345})+i(e_5 - e_{12345}) + c_2 (e_{15}+e_{2345}) + c_3 (e_{14}-e_{23})~,
\end{equation}
or
\begin{equation}
\label{nullx2}
\nu^2 = k_1 (e_5+e_{12345}) + e_{15}+e_{2345} + i k_2 (e_1 - e_{234})
+ k_3 (e_2 - e_{134})~,
\end{equation}
or
\begin{equation}
\label{nullx3}
\nu^2 = ix (1-e_{1234}) + e_5 + e_{12345}~,
\end{equation}
or
\begin{equation}
\label{nullx4}
\nu^2 = iy (1-e_{1234}) + \tau (e_1+e_{234})~,
\end{equation}
where $c_1, c_2, c_3, k_1, k_2, k_3, x, y, \tau$ are real functions.

Further simplification is possible. This is because if for one of the above normals the associated
1-form bi-linear is not null, then the corresponding case is not new but part of the cases for which
the first normal is $\nu^1=1+e_{12345}$. Thus the new cases which arise for $\nu^1=1+e_{1234}$ are those
for which both normal spinors are associated with null 1-form bi-linears. Evaluating the norm of the 1-form bi-linears
for  ({\ref{nullx1}})-({\ref{nullx4}}),
one finds $-16 c_3^2 (1+c_1^2)$, $-16(k_2^2+k_3^2)$, $-16x^2$ and $0$ respectively.
Setting these expressions to zero, one obtains the solutions $c_3=0$, $k_2=k_3=0$ and $x=0$ in
({\ref{nullx1}})-({\ref{nullx3}}). Using this, the cases ({\ref{nullx1}})-({\ref{nullx3}}) can be combined as
\begin{equation}
\label{nully1}
\nu^2 = b_1 (e_5 + e_{12345}) + i b_2 (e_5-e_{12345}) + b_3 (e_{15}+e_{2345})
\end{equation}
where  $b_1, b_2, b_3$ are real functions. In fact, an additional simplification is possible  by requiring
that the 1-form bi-linear associated with $\nu^2+\nu^1$ be null which forces $b_2=0$. This is because the second normal
must be linearly independent and can be defined
up to choice of the first one.

To summarize, when $\nu^1=1+e_{1234}$,  one can without loss of generality choose
\begin{equation}
\label{nullt1}
\nu^2 = a(e_5 + e_{12345}) + b (e_{15}+e_{2345})
\end{equation}
or
\begin{equation}
\label{nullt2}
\nu^2 = i m (1-e_{1234}) + n (e_1+e_{234})
\end{equation}
where  $a, b, m, n$ are real functions.

\section{Solution of ${\cal R}$-identities}

Having specified the normals, the Killing spinors are determined using the orthogonality condition. This allows us to express the $T$
components of supercurvature in terms of  $u$'s. Substituting this into the ${\cal R}$-identities, one obtains linear conditions on the
$u$'s. In many cases, the linear conditions on the $u$'s imply that all the $u$'s vanish and so
such backgrounds are locally maximally supersymmetric. In some other cases, the linear system for the $u$'s does {\it not} imply that
all the $u$'s vanish. As a result it may appear that there could be some non-trivial solutions. However, after taking into account the
{\it explicit dependence} of $T$'s in terms of the physical fields, one finds that all the $u$'s are forced to vanish.

\subsection{$Spin(7)\ltimes{\mathbb{R}}^9$}

We have shown that if the first normal is  $\nu^1=1+e_{1234}$, then
$\nu^2$ can be chosen either as ({\ref{nullt1}}) or as ({\ref{nullt2}}). Therefore there are two cases to investigate
which in turn can be separated into different subcases.

\subsubsection{$\nu^1=1+e_{1234}$, $\nu^2 = a (e_5+e_{12345})+b (e_{15}+e_{2345})$}

To proceed, we solve the ${\cal R}$-identities for the $u$'s first in the special cases for which either  $b$ or $a$ vanishes,
and then for  the case  $a,b\not=0$.
If $b \neq 0$, then after a computer assisted computation, one finds that the linear system implies that $u=0$,
and hence the solutions are locally maximally supersymmetric.

In the remaining case, for which $\nu_2=e_5+e_{12345}$, one finds that after solving the
${\cal{R}}$-identities, there is one real u degree of freedom remaining. In addition, none
of the $T^i$ vanish.

\subsubsection{$\nu^1=1+e_{1234}$, $\nu^2 = im (1-e_{1234}) + n (e_1+e_{234})$}
\label{nn}

This case is separated into various special cases. The ${\cal R}$-identities are solved for all these
and it turns out that some of the $u$'s do not vanish. In particular, we find the following.

\begin{itemize}

\item[(i)] If $\nu^2 = i (1-e_{1234})$, the ${\cal R}$-identities  are not  sufficient to set all $u$'s to zero. In fact after solving
 the ${\cal R}$-identities,
one finds that there are 78 real $u$ degrees of freedom remaining. Nevertheless substituting the solution
of the ${\cal R}$-identities into (\ref{supcurv2}), one finds that
\begin{equation}
T^1=0, \qquad T^2=0 ~,
\end{equation}
 However although several components of $T^3$, $T^4$ and $T^5$ vanish,
$T^3$, $T^4$ and $T^5$ are not zero.

\item[(ii)] If $\nu^2=e_1+e_{234}$, the ${\cal R}$-identities imply that all, but 3 real $u$ degrees of freedom, vanish.
Substituting this result into (\ref{supcurv2}), one finds that
\begin{equation}
T^1=0, \qquad T^3=0 ~,
\end{equation}
 However,
$T^2$, $T^4$ and $T^5$ are not necessarily zero.

\item[(iii)] If both $n,m$ are non-vanishing and so the direction of the second normal can be chosen as $
\nu^2 = i (1-e_{1234}) +y (e_1+e_{234})$, the ${\cal R}$-identities again imply that
 3 real $u$ degrees of freedom remain. Again one finds that
\begin{equation}
T^1=0 ~,
\end{equation}
however
$T^2$, $T^3$, $T^4$ and $T^5 $ are not necessarily zero.

\end{itemize}

\subsection{$SU(5)$}

We have shown that if the first normal is $\nu^1=1+e_{12345}$, there are distinct choices for the second normal
given in (\ref{snorm1}) and in (\ref{snorm2}). In the solution of the ${\cal R}$-identities these in turn separate
into different subcases depending on the non-vanishing components of the second normal.

\subsubsection{$\nu^1=1+e_{12345}$, $\nu^2 = ix (1-e_{12345}) + \sigma^{12}(e_{12}-e_{345}) + \sigma^{34}(e_{34}-e_{125})$}
\label{tt}

To investigate the various subcases observe that if one $\sigma$'s is non-vanishing, then without loss of generality
we can choose it to be $\sigma^{12}$. This is because the orbits represented by $(e_{12}-e_{345})$ and $(e_{34}-e_{125})$
can be treated symmetrically-they are interchanged by the lexicographic transformation $12\leftrightarrow 34$.
Thus from now on, in such case, we shall choose the normal direction by setting $\sigma^{12}=1$. We also write  $\sigma^{34}=\beta$.

The various subcases that arise are as follows.

\begin{itemize}

\item[(i)] If both $\sigma$ components vanish and so $
\nu^2 = i (1-e_{12345})$, the ${\cal R}$-identities imply that  78 real $u$ degrees of freedom remaining.
Nevertheless,  one finds that
\begin{equation}
T^1=0, \qquad T^2=0 ~.
\end{equation}
 In addition,
several components of $T^3$, $T^4$ and $T^5$ vanish. However, the ${\cal R}$-identities do not force $T^3$, $T^4$ and $T^5$ to vanish.

\item[(ii)] If $\beta \neq 0$, then  $u=0$ and so ${\cal R}=0$. Therefore all such backgrounds
 are locally maximally supersymmetric.

\item[(iii)] If $x \neq 0$, $\beta=0$, the ${\cal R}$-identities imply that all, but 2 real $u$ degrees of freedom, vanish. Moreover, one can show
that
\begin{equation}
T^1=0, \qquad T^3=0, \qquad T^4=0~.
\end{equation}
 In addition,
several components of $T^2$ and $T^5$ vanish. However, the ${\cal R}$-identities do not force $T^2$ and $T^5$ to vanish.

\item[(iv)] If $x=\beta =0$, the ${\cal R}$-identities imply again that  2 real $u$ degrees of freedom are not vanishing.
In case (iii) above
\begin{equation}
T^1=0, \qquad T^3=0, \qquad T^4=0~.
\end{equation}
 However, although several components of $T^2$ and $T^5$ vanish,
$T^2 \neq 0$ and $T^5 \neq 0$.

\end{itemize}

\subsubsection{$\nu^1=1+e_{12345}$,
$\nu^2 =  ix (1-e_{12345}) + e_1 +e_{2345}
+ \sigma^{12} (e_{12}-e_{345}) + \sigma^{34} (e_{34}-e_{125}) + \sigma^{45}(e_{45}-e_{123})
+\sigma^{23} e_{23} - {\bar{\sigma}}^{23} e_{145}$}

This case can be separated into various subcases depending on the non-vanishing components of the second normal.
In all the subcases that arise, the ${\cal R}$-identities imply that $u=0$ and so ${\cal R}=0$. Thus all these backgrounds
are locally maximally supersymmetric.

\section{Local maximal supersymmetry}

Having solved the ${\cal R}$-identities, we have found that in a number of cases some of the $u$'s do not vanish.
To make further progress, we shall utilize the explicit dependence of the $T$'s in terms of the physical fields. As we shall show, the resulting additional conditions are  sufficient to show that all $T$'s vanish, and so all backgrounds with 30 supersymmetries are locally maximally supersymmetric.

\subsection{Solutions with $T^1=T^3=T^4=0$}

These $T$'s vanish in  the cases (iii) and (iv) of \ref{tt}.
To solve these conditions, we first observe that
 $T^1=0$ implies that
\begin{eqnarray}
F\wedge F=0~,
\end{eqnarray}
which in turn gives
\begin{eqnarray}
i_XF\wedge F=0~.
\end{eqnarray}
Substituting this into $T^3=0$ and using the Bianchi $dF=0$, one finds that
\begin{eqnarray}
\nabla F=0~,
\label{covconst}
\end{eqnarray}
i.e. $F$ is covariantly constant with respect to the Levi-Civita connection $\nabla$.

It remains to explore $T^4=0$. For this observe that if $T^4=0$ then ({\ref{algcon3}})
implies that
\begin{eqnarray}
F_{C[A_1A_2A_3} F^C{}_{ A_4]MN}=0~.
\end{eqnarray}
This is the fundamental identity of a Lorentzian 3-Lie algebra. The solutions of this identity   have been classified in \cite{jose}.
Applying the classification results to our case, we find that the solutions for $F$ are
either
\begin{eqnarray}
F=\lambda_1 d{\rm Vol}(V_1)+\lambda_2 d{\rm Vol}(V_2)~,
\label{sol1}
\end{eqnarray}
where $\lambda_1, \lambda_2$ are constants and $V_1$ and $V_2$ are orthogonal  4-planes such that at most one of them  is Lorentzian and the rest Euclidean;
or
there is a null 1-form $v$ such that
\begin{eqnarray}
F=v\wedge \varphi~,~~~~
\label{sol2}
\end{eqnarray}
and $\varphi$ are the structure constants of a Euclidean metric Lie algebra, $\mathfrak{g}$;   or
\begin{eqnarray}
F=v\wedge \varphi+ \lambda d{\rm Vol}(V)
\label{sol3}
\end{eqnarray}
where $V$ is a Euclidean 4-plane orthogonal to the Lie algebra $\mathfrak{g}$. Since ${\rm dim}\, \mathfrak{g}\leq 9$, the semisimple
Lie algebras  that may occur
are
\begin{eqnarray}
\mathfrak{su}(2)~,~~~\mathfrak{su}(2)\oplus \mathfrak{su}(2)~,~~~~ \mathfrak{su}(2)\oplus \mathfrak{su}(2) \oplus \mathfrak{su}(2), \ \
\mathfrak{su}(3)~.
\end{eqnarray}

However, $F\wedge F=0$. For the solution (\ref{sol1}) this implies that $\lambda_1 \lambda_2=0$ and so either
$\lambda_1=0$ or $\lambda_2=0$. In either case
\begin{eqnarray}
F=\lambda d{\rm Vol}(V)
\label{sol4}
\end{eqnarray}
is a simple form, but there are two cases to consider depending on whether $V$ is a Euclidean or a Lorentzian plane. The solution (\ref{sol2})
satisfies $F\wedge F=0$ automatically. Applying $F\wedge F=0$ in (\ref{sol3}) and assuming that $\varphi\not=0$, one concludes that $\lambda=0$.
As a result, the solution of the conditions which arise from $T^1=T^3=T^4=0$ implies that either $F$ is simple and it is given in
 (\ref{sol4}) for $V$ a Euclidean or a Lorentzian 4-plane, or $F$ is given in  (\ref{sol2}).

It remains to examine whether  $T^2$ and $T^5$ vanish. It turns out that it suffices to show that $T^5=0$ since in all cases under consideration
in this section a direct inspection of $T^2$ and $T^5$ implies that if $T^5$ vanishes so does $T^2$. Moreover $T^5$ can be simplified as
\begin{eqnarray}
(T^5_{MN})_{A_1\dots A_5}&=&\frac{1}{(72)^2 } \big[-6 F_{M B_1B_2 B_3} F_{N C_1C_2 C_3}
  \epsilon^{B_1B_2 B_3C_1C_2 C_3}{}_{A_1\dots A_5}
  \cr
  &&
  +9F_{LPB_1B_2} F^{LP}{}_{C_1C_2 }
  \epsilon_{MN}{}^{B_1B_2C_1C_2 }{}_{A_1\dots A_5}\big]
  ~.
  \label{simt5}
  \end{eqnarray}

Now if $F$ is simple and so given in (\ref{sol4}),  $T^5=0$. Thus $T^2=0$ and so all  such solutions are
 locally maximally supersymmetric. Hence, the only  remaining possibility is that for which $F$
is given by  (\ref{sol2}).

To proceed, observe that if $F$ is given by (\ref{sol2}), then the second term
in (\ref{simt5}) vanishes.
If a  solution exists and $F$ is given as in (\ref{sol2}), the null vector field associated with $v$, also denoted by $v$,
satisfies
\begin{equation}
\label{contractv1}
v^M  (S_{M Q_1})_{ Q_2 Q_3 Q_4 Q_5 Q_6 Q_7}= v^M (S_{Q_1 Q_2})_{ M Q_3 Q_4 Q_5 Q_6 Q_7}=0~,
\end{equation}
where
\begin{equation}
(S_{N_1 N_2})_{M_1 M_2 M_3 M_4 M_5 M_6} = F_{N_1 [M_1 M_2 M_3} F_{|N_2| M_4 M_5 M_6]} \ .
\end{equation}
It is straightforward to verify, by direct computation, that in cases $(iii)$ and $(iv)$ of section \ref{tt},
if $T^5 \neq 0$, there are no null vector fields satisfying ({\ref{contractv1}}). Hence these cases
must in fact have $T^5=0$, and hence be locally maximally supersymmetric.

\subsection{Analysis of the Remaining Solutions}

The remaining solutions consist of the $Spin(7)$ cases with $\nu^2=e_5+e_{12345}$, $\nu^2=e_1+e_{234}$,
$\nu^2=i(1-e_{1234})+y(e_1+e_{234})$ ($y \in {\mathbb{R}}$, $y \neq 0$), and $\nu^2=i(1-e_{1234})$. There is also
a $SU(5)$ solution with $\nu^2=i(1-e_{12345})$. The analysis of these solutions is somewhat more involved, and
the details are presented in Appendices B and C. In all cases, one finds that the solutions are locally maximally
supersymmetric.

 \section{Discrete Quotients}

 So far, we have ruled out the existence of local geometries that preserve 30 supersymmetries in 11-dimensional
supergravity. To prove that there are no solutions that preserve 30 supersymmetries, it remains to show that there
are no such backgrounds which can be constructed as discrete quotients of maximally supersymmetric ones. The
simply connected maximally supersymmetric backgrounds are isometric \cite{georgejose} to Minkowski ${\mathbb{R}}^{10,1}$,
Freund-Rubin $AdS_4\times S^7$ and $AdS_7\times S^4$ \cite{fr},
and plane wave \cite{kg} $CW_{11}$ solutions. New backgrounds that preserve less than maximal supersymmetry can arise by taking
appropriate  quotients  of these backgrounds with discrete subgroups of their isometry groups.  The general
procedure for investigating the number of supersymmetries preserved by such discrete quotients has been explained in \cite{josesimon}. It has also
 been applied in \cite{josegadhia} to rule out the existence of discrete quotients with 31 supersymmetries in
11-dimensional supergravity, and in \cite{28iib, more28iib} to rule out the existence of such backgrounds with 28 and 30
supersymmetries in IIB supergravity. Because the general method has already been explained in detail, we shall
not elaborate apart from saying that it suffices to consider   elements in the appropriate
isometry groups which lie in the image of the exponential map, ie they are written as $e^X$ where $X$ is an element
of the Lie algebra of the isometry group. Moreover $X$ can be specified up to a conjugation. As a result,
 $X$ can be put onto a maximal torus. Since the isometry groups are Lorentzian
 there are different maximal tori and so different canonical forms for $X$ leading to several different cases that should be investigated.
 We shall apply this general procedure for the Minkowski and plane wave backgrounds. It turns out that for the AdS backgrounds
 a simpler argument can be used to rule out the existence of $N=30$ backgrounds.

 \subsection{Minkowski}

 The isometry group of Minkowski space is the Poincar\'e group $SO(10,1)\ltimes {\mathbb{R}}^{10,1}$. It is easy to see that identifications
 along the subgroup of translations preserve all supersymmetry. Thus to preserve less than maximal supersymmetry, one should consider
 discrete subgroups of the Lorentz group. Suppose that $X\in \mathfrak{spin}(10,1)$. Up to a conjugation, $X$ can be written either as
 \begin{eqnarray}
 X={1\over2} [\theta_0 \Gamma_{05}+\theta_1 \Gamma_{16}+\theta_2\Gamma_{27}+\theta_3 \Gamma_{38}+\theta_4 \Gamma_{49}]~,
 \label{fcase}
 \end{eqnarray}
 or as
 \begin{eqnarray}
X={1\over2} [\theta_1 \Gamma_{16}+\theta_2\Gamma_{27}+\theta_3 \Gamma_{38}+\theta_4 \Gamma_{49}+\theta_5\Gamma_{5{\natural}}]~,
\label{scase}
\end{eqnarray}
or as
\begin{eqnarray}
 X={1\over2} [\sqrt{2} \Gamma_{\natural} (\Gamma_0+\Gamma_5) +\theta_1 \Gamma_{16}+\theta_2\Gamma_{27}+\theta_3 \Gamma_{38}+\theta_4 \Gamma_{49}]~.
 \label{tcase}
 \end{eqnarray}

Let us first consider the (\ref{fcase}) case first. Decompose the spinor representations $\Delta_{32}$ of $Spin(10,1)$ in representations
of the commuting elements $\Gamma_{05}, \Gamma_{16}, \Gamma_{27}, \Gamma_{38}$ and $ \Gamma_{49}$. One finds that  $\Delta_{32}=\oplus_{(\sigma_0, \dots, \sigma_4)}W_{\sigma_0\sigma_1\dots \sigma_4}$
and $X$ becomes
\begin{eqnarray}
 X={1\over2} [\theta_0 \sigma_0+i\theta_1 \sigma_1+i\theta_2\sigma_2+i\theta_3 \sigma_3+i\theta_4 \sigma_4]~,
\end{eqnarray}
where $\sigma_0, \dots, \sigma=\pm 1$.

Now  assume that $e^X$ preserves 30 spinors. In such case,  there is a  choice of $\sigma$'s such that both  $W_{\sigma_0\sigma_1\dots \sigma_4}$ and
 $W_{\sigma_0\bar\sigma_1\dots \bar\sigma_4}$ with $\bar\sigma=-\sigma$ are invariant, ie $e^X=1$ for both cases. Using this, one concludes
that
\begin{eqnarray}
e^{\sigma\theta_0}=1
\end{eqnarray}
and so $\theta_0=0$. Supersymmetry is not preserved under time-like identifications as expected.

Using this next observe that if for some $\sigma$'s $W_{\sigma_0\sigma_1\dots \sigma_4}$ is invariant, then the subspaces
$W_{+\sigma_1\dots \sigma_4}$, $W_{-\sigma_1\dots \sigma_4}$, $W_{+\bar\sigma_1\dots \bar\sigma_4}$ and
$W_{-\bar\sigma_1\dots \bar\sigma_4}$ are also invariant. Therefore the invariant subspaces have dimension $4k$ and so
backgrounds with 30 supersymmetries cannot arise this way.

To investigate the second case (\ref{scase}), again decompose the spinor representation $\Delta_{32}$ in eigenspaces $W_{\sigma_1\dots\sigma_5}$
 of $\Gamma_{ii+5}$, $i=1,2,3,4,5$,
 and write $X$ as
\begin{eqnarray}
 X={1\over2} [i\theta_1 \sigma_1+i\theta_2\sigma_2+i\theta_3 \sigma_3+i\theta_4 \sigma_4+i\sigma_5\theta_5]~,
\end{eqnarray}
where $\sigma_1,\dots,\sigma_5=\pm1$.
Now in order the discrete elements to preserve precisely 30 supersymmetries, the invariant subspaces should be in complex conjugate pairs. As a result
the non-invariant subspace should be the sum of a 1-dimensional subspace and its complex conjugate. Without loss of generality, assume
that the non invariant subspace is $W_{1,1,1,1,1}\oplus W_{-1,-1,-1,-1,-1}$. Since $e^X=1$ for $\sigma_1=-1, \sigma_2=\sigma_3=\dots=\sigma_5=1$ and
$\sigma_1=1, \sigma_2=-1, \sigma_3=\dots=\sigma_5=1$ multiplying the two expressions of $e^X$ together, we find that
\begin{eqnarray}
e^{i[\theta_3+\theta_4+\theta_5]}=1~.
\label{345}
\end{eqnarray}
Next, multiply both sides of $e^X=1$ for $\sigma_1=\sigma_2=1, \sigma_3=\sigma_4=\sigma_5=-1$
with ({\ref{345}}).
One concludes that $e^X=1$ for $\sigma_1=\dots=\sigma_5=1$, and so $W_{1,1,1,1,1}\oplus W_{-1,-1,-1,-1,-1}$ is also invariant. Therefore assuming that
30 supersymmetries are preserved, one finds that all 32 of the supersymmetries are preserved and so there are no backgrounds
with 30 supersymmetries which can arise as discrete quotients in this way.

It remains to investigate the null case (\ref{tcase}). $e^X$ can be written as
\begin{eqnarray}
e^X=e^R (1+\Gamma_{\natural} \Gamma_+)
\end{eqnarray}
where
\begin{eqnarray}
R={1\over2}[\theta_1 \Gamma_{16}+\theta_2\Gamma_{27}+\theta_3 \Gamma_{38}+\theta_4 \Gamma_{49}]~.
\end{eqnarray}
Decomposing the spinor as $\epsilon=\epsilon_++\epsilon_-$, with $\Gamma_+\epsilon_+=0$, ie $\Delta_{32}=W_+\oplus W_-$,  one has that the invariance equations can be rewritten as
\begin{eqnarray}
\label{concc1}
e^R \epsilon_++ e^R \Gamma_{\natural} \Gamma_+\epsilon_-=\epsilon_+
\cr
e^R\epsilon_-=\epsilon_-
\end{eqnarray}
To preserve 30 supersymmetries, either $e^R$ leaves invariant either all of $W_-$, or a co-dimension one
or codimension two subspace $I$ in $W_-$.
When $e^R$ leaves all of $W_-$ invariant, ({\ref{concc1}}) implies that $e^R=1$,
and so the first equation implies that all $\epsilon_-$ must vanish.
The background preserves 1/2 of supersymmetry.

If $e^R$ does not leave the whole of $W_-$ invariant, decompose
$W_-$ into representations
of the commuting elements $\Gamma_{16}, \Gamma_{27}, \Gamma_{38}$ and $ \Gamma_{49}$;
$W_-=\oplus_{(\sigma_1 \dots \sigma_4)}Z_{\sigma_1\dots \sigma_4}$. Observe that
if $Z_{\sigma_1\dots \sigma_4}$ is invariant under $e^R$ then so is
$Z_{\bar\sigma_1\dots \bar\sigma_4}$, where $\bar\sigma=-\sigma$. As the invariant subspaces occur in complex conjugate
pairs, it follows that there cannot be a co-dimension 1 subspace $I \subset W_-$ invariant under $e^R$.
One can also exclude the possibility of a co-dimension 2 invariant subspace of $W_-$ by taking,
without loss of generality, the non-invariant subspace to be $Z_{+1+1+1+1} \oplus Z_{-1-1-1-1}$.
Then as $e^R=1$ on $W_-$ for $\sigma_1=1, \sigma_2=-1, \sigma_3=1, \sigma_4=1$
and also for $\sigma_1=-1, \sigma_2=1, \sigma_3=1, \sigma_4=1$, multiplying the two expressions for
$e^R$ together gives
\begin{equation}
\label{334}
e^{i(\theta_3+\theta_4)}=1 \ .
\end{equation}
Next, multiply both sides of $e^R=1$ for $\sigma_1=\sigma_2=1, \sigma_3=\sigma_4=-1$
with ({\ref{334}}); one finds that $Z_{+1+1+1+1} \oplus Z_{-1-1-1-1}$ must also be invariant.

In conclusion, there are no discrete
quotients of Minkowski space ${\mathbb{R}}^{10,1}$ which preserve 30 supersymmetries.

\subsection{$AdS_4\times S^7$ and $AdS_7\times S^4$}

The spinor $\Delta_{32}$ representation of $Spin(10,1)$ is decomposed under the
isometry group $SO(3,2)\times SO(8)$ of $AdS_4\times S^7$  as $\Delta_4\times \Delta_8^+$, where $\Delta_4$ is the Majorana representation
of $Spin(3,2)$ and $\Delta_8^+$ is the Majorana-Weyl representation of $Spin(8)$. Invariant subspaces of discrete subgroups
of the isometry groups have dimension $nm$, where $n\leq 4$ and $m\leq 8$. Since 30 cannot be written this way, there are no discrete quotients
of the $AdS_4\times S^7$ background which preserve 30
supersymmetries.

Similarly $\Delta_{32}$ representation of $Spin(10,1)$ is decomposed under the
isometry group $SO(6,2)\times SO(5)$ of $AdS_7\times S^4$  as $\Delta_8^+\times \Delta_4$, where $\Delta^+_8$ is the Weyl representation
of $Spin(6,2)$ and $\Delta_4$ is the Dirac representation of $Spin(8)$. Again the dimension of the invariant subspaces should be $nm$
and so there are no discrete quotients
of the  $AdS_7\times S^4$ background which preserve   30 supersymmetries.

\subsection{Plane wave}

The symmetry superalgebra of the maximally supersymmetric plane wave solution \cite{kg} of 11-dimensional supergravity
has been computed in \cite{fofgp}. The investigation of the existence of discrete quotients of the plane wave solution
which preserve 30 supersymmetries is similar to that done  in \cite{josegadhia} for the existence of discrete
quotients that preserve 31 supersymmetries. However, there are some differences because the requirement
of 30 supersymmetries is weaker. Because of this, we shall repeat some of the steps of the analysis.

To examine the supersymmetry preserved by the  discrete quotients of the maximally supersymmetric plane wave, one
needs the bosonic part of the symmetry superalgebra and the way that the bosonic generators act on the spinorial generators.   The bosonic
part of the superalgebra has generators $(e_+, e_-, e_i, e_i^*)$ and $(M_{ij})$ for $i,j\leq 3$ and $i,j\geq 4$, $i,j=1,\dots,9$.
The commutators of the bosonic generators are
\begin{eqnarray}
&&[e_-, e_i]=e^*_i~,~~~[e_-, e^*_i]=-{\mu^2\over 9} e_i~~~(i\leq 3)~,~~~[e_-, e^*_i]=-{\mu^2\over36} e_-~~~(i\geq 4)~,
\cr
&&[e^*_i, e_j]=-{\mu^2\over9} \delta_{ij} e_+~~~(i,j\leq 3)~,~~~[e^*_i, e_j]=-{\mu^2\over36} \delta_{ij} e_+~~~(i,j\geq 4)~,
\cr
&& [M_{ij}, M_{kl}]=-\delta_{ik} M_{jl}+ \delta_{jk} M_{il}-(k\leftrightarrow l)~~(i,j,k,l\leq 3)~{\rm and}~ (i,j,k,l\geq 4)~,
\cr
&&[M_{ij}, e_k]=-\delta_{ik} e_j+\delta_{jk} e_i~,~~~[M_{ij}, e^*_k]=-\delta_{ik} e^*_j+\delta_{jk} e^*_i~.
\end{eqnarray}
In particular the generators $(M_{ij})$ span the Lie algebra $\mathfrak{so}(3)\oplus \mathfrak{so}(6)$.
The commutators of the bosonic generators with  the spinorial generators $Q_\pm$ are
\begin{eqnarray}
&&[e_+, Q_\pm]=0~,~~~[e_-, Q_+]=-{\mu\over4} IQ_+~,~~~[e_-, Q_-]=-{\mu\over12} IQ_-
\cr
&&[e_i, Q_+]=-{\mu\over6} I\Gamma_i \Gamma_+ Q_-~~~(i\leq 3)~,~~~[e_i, Q_+]=-{\mu\over12} I\Gamma_i \Gamma_+ Q_-~~~(i\geq 4)~,
\cr
&&[e^*_i, Q_+]=-{\mu^2\over18} \Gamma_i \Gamma_+ Q_-~~~(i\leq 3)~,~~~[e^*_i, Q_+]=-{\mu^2\over72} \Gamma_i \Gamma_+ Q_-~~~(i\geq 4)~,
\cr
&&[M_{ij}, Q_\pm]={1\over2} \Gamma_{ij} Q_\pm~,~~~(i,j\leq 3)~{\rm and}~ (i,j\geq 4)~,
\end{eqnarray}
where $I=\Gamma_{123}$ and $\Gamma_\pm Q_\pm=0$.

The most general Lie algebra element of the symmetry group of the background is
\begin{eqnarray}
X= u^- e_-+u^+ e_++v^i e_i+w^i e_i^*+\theta_1M_{12}+\theta_2M_{45}+ \theta_3 M_{67}+\theta_4 M_{89}
\end{eqnarray}
where we have used the conjugation by $SO(3)\times SO(6)$ to put the component of $X$ long $\mathfrak{so}(3)\oplus \mathfrak{so}(6)$
in the Cartan subalgebra.
Inspecting the commutators of the bosonic generators with the spinorial ones, $X$ acts on the spinors as
\begin{eqnarray}
X&=&-[{\mu\over4} I \Pi_++{\mu\over12} \Pi_-] u^--\sum_i{\lambda_i\over2} v^i I\Gamma_i\Gamma_+- \sum_i{\lambda^2_i\over2} w^i \Gamma_i\Gamma_+
\cr
&&+
{1\over2}\theta_1\Gamma_{12}+{1\over2}\theta_2\Gamma_{45}+ {1\over2}\theta_3 \Gamma_{67}+{1\over2}\theta_4 \Gamma_{89}~,
\label{pplift}
\end{eqnarray}
where $\Pi_\pm$ are projections, $\Pi_\pm Q_\pm=Q_\pm$, $\Pi_+^2=\Pi_+$, $\Pi_-^2=\Pi_-$, $\Pi_++\Pi_-={\bf 1}$ and $\Pi_+\Pi_-=\Pi_-\Pi_+=0$,
and $\lambda_i={\mu\over 3}$ for $i\leq 3$ and $\lambda_i={\mu\over 6}$ for $i\geq 4$.
Since (\ref{pplift}) does not depend on $u^+$, any identification along this direction will preserve all the supersymmetry of the background. Furthermore
decomposing the spinor representation  as $W_+\oplus W_-$, where $\Gamma_+ W_+=\Gamma_-W_-=0$, ie $\epsilon_\pm =\Pi_\pm \epsilon$,
the
invariance  condition $e^X\epsilon=\epsilon$ can be written as
\begin{eqnarray}
e^A \epsilon_++ \Gamma_+ \beta \epsilon_-=\epsilon_+~,
\cr
e^A \epsilon_-=\epsilon_-~,
\label{splitpp}
\end{eqnarray}
 where
\begin{eqnarray}
A=-[{\mu\over4} I \Pi_++{\mu\over12}  I \Pi_-] u^-+
{1\over2}\theta_1\Gamma_{12}+{1\over2}\theta_2\Gamma_{45}+ {1\over2}\theta_3 \Gamma_{67}+{1\over2}\theta_4 \Gamma_{89}~,
\end{eqnarray}
and $\beta$ is an involved expression\footnote{In \cite{josegadhia}, $\beta$ is denoted with $\check\alpha$.} associated with the components of $X$ that contain $\Gamma_+$ which
its precise form is not needed at present.

To continue, first observe that for the invariance condition  on $\epsilon_-$, $e^A\epsilon_-=\epsilon_-$ $A$ is simplified as
\begin{eqnarray}
A={-{\mu\over12} I u^-+{1\over2}\theta_1\Gamma_{12}+{1\over2}\theta_2\Gamma_{45}+ {1\over2}\theta_3 \Gamma_{67}+{1\over2}\theta_4 \Gamma_{89}}
\end{eqnarray}
To preserve 30 supersymmetries,  $e^A$ should  leave invariant either  all $W_-$ or a codimension 2 subspace $S$. (As we shall see the codimension
1 case does not occur.) First we consider the
latter case to show if a codimension 2 subspace is invariant, then all $W_-$ is invariant. Since the generators $I$, $\Gamma_{12}$, $\Gamma_{45}$, $\Gamma_{67}$ and $\Gamma_{89}$ commute and square to $-{\bf 1}$, (the complexified)
$W_-$ can be decomposed
in their eigenspaces as $W_-=\oplus_{\sigma_0\dots \sigma_4} W_{\sigma_0\dots \sigma_4}$, where $\sigma_0, \dots\sigma_4=\pm1$ and
$\sigma_0\sigma_2\sigma_3\sigma_4=-1$. Then
\begin{eqnarray}
A={-{\mu\over12}iu^-\sigma_0 +{i\over2}\theta_1\sigma_1+{i\over2}\theta_2\sigma_2+ {i\over2}\theta_3 \sigma_3+{i\over2}\theta_4 \sigma_4}~.
\label{minusa}
\end{eqnarray}
Clearly if for some choice of $\sigma$'s, $W_{\sigma_0\dots \sigma_4}$ is  invariant, then the complex conjugate subspace
$W_{\bar\sigma_0\dots \bar\sigma_4}$, where $\bar\sigma=-\sigma$, is also invariant. Thus the invariant subspaces are always
of even codimension.

Next assume without loss of generality that $W_{+1+1+1+1-1}\oplus W_{-1-1-1-1+1}$ is not an invariant subspace and the remaining eigenspaces
are invariant. This implies that $e^A=1$ for $\sigma_0=\sigma_2=-\sigma_3=\sigma_4=\sigma_1=1$ and $e^A=1$ for
$-\sigma_0=-\sigma_2=\sigma_3=-\sigma_4=\sigma_1=1$. Using this, one concludes that $e^{i\theta_1}=1$. In addition
$e^A=1$ for $\sigma_0=\sigma_2=\sigma_3=-\sigma_4=-\sigma_1=1$. Multiplying $e^A$ with this choice of $\sigma$'s with $e^{i\theta_1}=1$, one finds that
$W_{+1+1+1+1-1}$ is also invariant. Therefore, if one assumes that a codimension 2 subspace of $W_-$ is invariant, then all
$W_-$ is invariant.

Assuming that all $W_-$ is invariant to make further progress, one has to examine the action of $e^A$ on $W_+$. Again $W_+$
can be decomposed as $W_+=\oplus_{\sigma_0\dots \sigma_4} W_{\sigma_0\dots \sigma_4}$ in eigenspaces of the  generators
$I$, $\Gamma_{12}$, $\Gamma_{45}$, $\Gamma_{67}$ and $\Gamma_{89}$ but now $\sigma_0\sigma_2\sigma_3\sigma_4=1$,
where $\sigma_0, \dots\sigma_4=\pm1$. Moreover
\begin{eqnarray}
A={-{\mu\over4}iu^-\sigma_0 +{i\over2}\theta_1\sigma_1+{i\over2}\theta_2\sigma_2+ {i\over2}\theta_3 \sigma_3+{i\over2}\theta_4 \sigma_4}~,
\label{plusa}
\end{eqnarray}
ie $e^A$ is represented differently on the $W_-$ and $W_+$ subspaces.
Using that $e^A=1$ for $A$ given in (\ref{minusa}) and taking into account that on $W_+$ $\sigma_0\sigma_2\sigma_3\sigma_4=1$, it is easy to show that
\begin{eqnarray}
e^A\epsilon_+=e^{-{\mu\over3}iu^-\sigma_0}\epsilon_+~.
\label{actplus}
\end{eqnarray}
However $e^A=1$ for $A$ given in (\ref{minusa}) implies that $\mu u^-=6n\pi$, $n\in {\mathbb{Z}}$. Substituting this into (\ref{actplus}),
one concludes that $e^A$ acts with the identity on $W_+$.  Thus the invariance condition (\ref{splitpp}) reduces to
\begin{eqnarray}
\Gamma_+\beta\epsilon_-=0~.
\end{eqnarray}
In order for a background to preserve 30 supersymmetries, the Kernel of $\beta$ should have dimension 14 for some choice of parameters
$u,w$. However it has been shown in \cite{josegadhia} that if $e^A=1$, the dimension of a non-trivial  Kernel is either 8 or 16. Thus
there are no discrete quotients of the maximally supersymmetric plane wave that preserve 30 supersymmetries.

\section{Plane waves and 28 supersymmetries}

It is clear that as in the case of IIB supergravity, the geometries of M-theory backgrounds with near maximal number of supersymmetries
are severely restricted.
It is natural to ask what is the highest possible fraction of supersymmetry, other than maximal,
that can be preserved. Although backgrounds with 29 supersymmetries cannot be ruled out,
the plane wave superalgebra construction of \cite{gunaydin} indicates that there may
be a plane wave solution that preserves 28 supersymmetries. This plane wave superalgebra is characterized by a
$(SO(3)\times SU(3)\times U(1))\ltimes H_9$ bosonic symmetry, where $H_9$ is the Heisenberg group with 19 generators.
 Assuming that this will be a symmetry of the background, one can analyze all plane wave solutions of M-theory with
 $(SO(3)\times SU(3)\times U(1))\ltimes H_9$ symmetry group. The most general plane wave ansatz with this symmetry is
 \begin{eqnarray}
 ds^2 = 2 dv (du + {1\over 2} \lambda_{ab} x^a x^b dv) + ds^2({\mathbb{R}}^9)~, ~~~F= dv \wedge \Phi \ ,
 \label{pw28}
\end{eqnarray}
where the transverse space ${\mathbb{R}}^9$ of the plane wave is decomposed as ${\mathbb{R}}^9={\mathbb{R}}^3\oplus {\mathbb{R}}^6$ under the
$SO(3)\times SU(3)$ symmetry, $ds^2({\mathbb{R}}^9)=(dx^a)^2$,
\begin{eqnarray}
\Phi&=&k\, d{\rm vol}({\mathbb{R}}^3)+ \mu \chi+\bar\mu \bar\chi~,~~~
\end{eqnarray}
where
$\chi$ is the $SU(3)$-invariant (3,0)-form on ${\mathbb{C}}^6$, and $(\lambda_{ab})=\lambda_1 {\bf 1}_{3\times 3}\oplus \lambda_2 {\bf 1}_{6\times 6}$.
The investigation of the Killing spinor equations is presented in appendix C. In particular,
one finds that
 such plane wave solutions preserve either 16, or 20, or 32 supersymmetries, depending on the choice of parameters
 $\lambda_1, \lambda_2, k, \mu$, but not 28. The solution with 20 supersymmetries has been found before in \cite{hg}.
 So we conclude that there is not  a plane wave solution with 28 supersymmetries
 and $(SO(3)\times SU(3)\times U(1))\ltimes H_9$ symmetry group. Of course, this does not rule out the existence of
 M-theory solutions with 28 supersymmetries. To establish the latter, an analysis  similar to that which has been
undertaken for IIB
 supergravity in \cite{28iib} is required.
 Nevertheless, it may turn out that the  nearly maximally supersymmetric backgrounds of   M-theory are more restricted than
 those of IIB because of the larger local Lorentz symmetry of the former.
At present, the highest number of supersymmetries known to be preserved by a non-maximally supersymmetric
solution is 26, for the case of  the plane wave solution found in  \cite{michelson}.
It is not known if this solution is the unique local solution with 26 supersymmetries, or if there are other
solutions with more supersymmetries than this.

\section{Concluding remarks}

We have shown that M-theory backgrounds that preserve 30 supersymmetries are maximally supersymmetric. First we have found that
all such backgrounds are locally maximally supersymmetric by demonstrating that the supercovariant curvature vanishes subject to field equations
and Bianchi identities, and then we proved that they cannot arise as discrete quotients of maximally supersymmetric ones. This result combined
with that of \cite{d11preon} for M-theory backgrounds with 31 supersymmetries leads to the conclusion that all M-theory
backgrounds with more than 29 supersymmetries are maximally supersymmetric. Moreover, we have explored the possibility of finding
a plane wave solution which preserves 28 supersymmetries with symmetry superalgebra that of \cite{gunaydin} which
has 28 odd generators and even subalgebra  $(\mathfrak{so}(3)\oplus \mathfrak{su}(3)\oplus \mathfrak{u}(1))\oplus_s
\mathfrak{H}_9$. We found that plane wave solutions with $(\mathfrak{so}(3)\oplus \mathfrak{su}(3)\oplus \mathfrak{u}(1))\oplus_s
\mathfrak{H}_9$ isometry algebra preserve either 16, 20 or 32 supersymmetries but not 28 depending on the choice of parameters.
The solution with 20 supersymmetries has been found before in \cite{hg}.

To classify nearly maximal supersymmetric solutions that preserve less than 30 supersymmetries, one can in principle repeat
the analysis we have done for the backgrounds with 30 supersymmetries. For example, the investigation of backgrounds
with 29 supersymmetries will require the choice of three linearly independent normal spinors and so on. It is clear that
for backgrounds with progressively less supersymmetry more normal spinors should be chosen, and so the gauge group will impose
less restriction on the choice of normals. The analysis will become increasingly involved. Nevertheless, it may be
possible to make further progress in constructing solutions with nearly maximal supersymmetry. This is based on the empirical  observation
that  if the normal spinors are chosen such that they have a large sigma group
\cite{sigmagp}, then the ${\cal R}$-identities impose
less restriction on the supercurvature ${\cal R}$. This increases the probability to find solutions which are not locally isometric
to maximally supersymmetric ones. An inspection of table 5 in \cite{inspin} suggests that there are five different possibilities
that can be explored for backgrounds with 28 supersymmetries in eleven dimensions. Although there is no guarantee that new solutions
will be found, it seems that these are the more promising cases to explore first.

\vskip 0.5cm
\noindent{\bf Acknowledgements} \vskip 0.1cm
We thank Murat Gunaydin for correspondence. UG is supported by the Swedish Research Council
and the Knut and Alice Wallenberg Foundation.
JG is supported by the EPSRC grant, EP/F069774/1. GP is partially supported
by the EPSRC grant, EP/F069774/1 and the STFC rolling grant ST/G000/395/1.
\vskip 0.5cm

\setcounter{section}{0}

\appendix{Normal spinors}
\label{appa}

\setcounter{subsection}{0}

In this section, we construct the generic normal spinors associated with
solutions of D=11 supergravity with a 30-dimensional space of Killing spinors.
For such solutions, the spinors are orthogonal (with respect to the $Spin(10,1)$
invariant inner product $B$) to two normal spinors $\nu^1$, $\nu^2$. Without
loss of generality, $\nu^1$, $\nu^2$ can be taken to be Majorana.
The conventions for the spinors, differential forms, gamma matrices and inner products are
identical to those in \cite{mtheor1, mtheor2}.

In particular, without loss of generality, the first normal spinor $\nu^1$ can be
written in a particularly simple form using $Spin(10,1)$ gauge transformations.
There are two possibilities, either $\nu^1$ is $SU(5)$ invariant, with
\begin{equation}
\nu^1 = 1+e_{12345}~,
\end{equation}
or $\nu^1$ is $Spin(7) \ltimes {\mathbb{R}}^9$ invariant with
\begin{equation}
\nu^1 = 1+e_{1234} \ ,
\end{equation}
where the two spinors have been expressed in the time-like and null spinor bases of \cite{mtheor1, mtheor2}, respectively.
In what follows, we shall consider these two cases separately.

\subsection{Solutions with $\nu^1 = 1+e_{12345}$}

For solutions with $SU(5)$ invariant $\nu^1$, it is particularly useful to work
in the timelike basis introduced in \cite{mtheor1}.
The generic form for the second Majorana normal is
\begin{eqnarray}
\nu^2 &=& \alpha \,1 + {\bar{\alpha}} e_{12345} + \beta^k e_k +{1 \over 4!} (\star {\bar{\beta}})^{m_1 m_2 m_3 m_4} e_{m_1 m_2 m_3 m_4}
\nonumber \\
&+& {1 \over 2} \sigma^{n_1 n_2} e_{n_1 n_2} -{1 \over 3!} (\star {\bar{\sigma}})^{k_1 k_2 k_3} e_{k_1 k_2 k_3}~,
\end{eqnarray}
where here $k,m,n=1, ... ,5$ and $\alpha, \beta^k, \sigma^{mn}$ are in general complex valued, and $\star$ denotes the Hodge
dual on ${\mathbb{R}}^5$.

There are two cases to consider depending on whether $\beta=0$ or $\beta \neq 0$.
Suppose that $\beta \neq 0$ and apply a $SU(5)$ gauge transformation to set
$\beta^2=\beta^3=\beta^4=\beta^5=0$, with $\beta^1=\beta$, and $\beta \in {\mathbb{R}}$.
Without loss of generality, set $\beta =1$.
Then apply a $SU(4)$ transformation in the $2,3,4,5$ directions to set
$\sigma^{13}=\sigma^{14}=\sigma^{15}=0$. Next, apply a $SU(3)$ transformation in
the $3,4,5$ directions to set $\sigma^{24}=\sigma^{25}=0$. Then apply a $SU(2)$
transformation in the $4,5$ direction to set $\sigma^{35}=0$ also.
Moreover  $\nu_2$ can be chosen up to $\nu_1$. Using this, the second normal can then
be written as
\begin{eqnarray}
\nu^2 &=& ix (1-e_{12345}) + e_1 +e_{2345} + \sigma^{12} e_{12} + \sigma^{23} e_{23}
+\sigma^{34} e_{34} + \sigma^{45} e_{45}
\nonumber \\
&-& {\bar{\sigma}}^{12} e_{345} - {\bar{\sigma}}^{23} e_{145} - {\bar{\sigma}}^{34} e_{125} - {\bar{\sigma}}^{45} e_{123}~,
\end{eqnarray}
for $x \in {\mathbb{R}}$.
Next, by applying a $SU(2)$ transformation in the $3,4$ directions, one can take
$\sigma^{45} \in {\mathbb{R}}$, and a $SU(2)$ transformation in the $4,5$ directions can be
used to set $\sigma^{34} \in {\mathbb{R}}$, and finally a $SU(4)$ transformation in the
$2,3,4,5$ directions can be used to set $\sigma^{12} \in {\mathbb{R}}$. The second normal then
simplifies to
\begin{eqnarray}
\nu^2 &=&  ix (1-e_{12345}) + e_1 +e_{2345}
+ \sigma^{12} (e_{12}-e_{345}) + \sigma^{34} (e_{34}-e_{125}) + \sigma^{45}(e_{45}-e_{123})
\nonumber \\
&+& \sigma^{23} e_{23} - {\bar{\sigma}}^{23} e_{145}~,
\end{eqnarray}
where $x, \sigma^{12}, \sigma^{34}, \sigma^{45} \in {\mathbb{R}}$.

In the second case, $\beta=0$, using the reasoning given in Appendix A of \cite{mtheor1},
one can apply a $SU(5)$ gauge transformation to write
\begin{equation}
\nu^2 = ix (1-e_{12345}) + \sigma^{12}(e_{12}-e_{345}) + \sigma^{34}(e_{34}-e_{125})~,
\end{equation}
for $x, \sigma^{12}, \sigma^{34} \in {\mathbb{R}}$.

\subsection{Solutions with $\nu^1 = 1+e_{1234}$}

For solutions with $Spin(7) \ltimes {\mathbb{R}}^9$ invariant $\nu^1$, it is particularly useful to work
in the null basis introduced in \cite{mtheor2}.
In this basis, the most general form for $\nu^2$ is

\begin{eqnarray}
\nu^2 &=& \alpha\,1 + {\bar{\alpha}} e_{1234} + w e_5 + {\bar{w}} e_{12345}
+ \tau^j e_j -{1 \over 3!} (\star {\bar{\tau}})^{n_1 n_2 n_3} e_{n_1 n_2 n_3}
\nonumber \\
&+& \psi^j e_{j5} -{1 \over 3!} (\star {\bar{\psi}})^{n_1 n_2 n_3} e_{n_1 n_2 n_3 5}
+{1 \over 2} (A^{ij} - (\star {\bar{A}})^{ij}) e_{ij} +{1 \over 2} (B^{ij} - (\star {\bar{B}})^{ij}) e_{ij5}~,
\nonumber \\
\end{eqnarray}
where here $i, j, n=1, ..., 4$, $\alpha, w, \tau^i, \psi^i, A^{ij}, B^{ij}$ are complex valued
and $\star$ denotes the Hodge dual on ${\mathbb{R}}^4$.

It is particularly useful to observe that under a ${\mathbb{R}}^9$ transformation generated by
$R^i \Gamma_{+i} + R^{\bar{i}} \Gamma_{+ \bar{i}} + \xi \Gamma_{+ \sharp}$ where
$R^{\bar{i}} = \overline{(R^i})$ and $\xi \in {\mathbb{R}}$, $w$, $\psi^i$ and $B^{ij}$ do not transform,
and
\begin{eqnarray}
\alpha &\rightarrow& \alpha +2 R_i \psi^i  + \sqrt{2} \xi w~,
\nonumber \\
\tau^j &\rightarrow& \tau^j -2 w R^j -2 R_i (B^{ij} - (\star {\bar{B}})^{ij}) +\sqrt{2} \xi \psi^j~,
\nonumber \\
A^{ij} &\rightarrow& A^{ij} + 4 R^{[i} \psi^{j]} + \sqrt{2} \xi B^{ij}~,
\end{eqnarray}
where here $R_i = \delta_{i \bar{j}} R^{\bar{j}}$.

To proceed, note that one can without loss of generality set $B^{ij}=0$ for all $i,j$.
To see this, first apply a $SU(3)$ transformation in the directions $1,2,3$
to set the coefficients of $e_{145}$ and $e_{245}$
to zero. Then
\begin{equation}
\label{auxspin}
w e_5 + {\bar{w}} e_{12345}  +{1 \over 2} (B^{ij} - (\star {\bar{B}})^{ij}) e_{ij5}
= w e_5 +  {\bar{w}} e_{12345} + \lambda e_{125} - \bar{\lambda} e_{345}
\end{equation}

Next consider the transformation generated by
\begin{equation}
X = {1 \over 2} \rho \big[e^{i \theta} \Gamma_{12} + e^{-i \theta} \Gamma_{\bar{1} \bar{2}}
+ e^{-i \theta} \Gamma_{34} + e^{i \theta} \Gamma_{\bar{3} \bar{4}} \big] \in spin(7)
\end{equation}
for $\rho, \theta \in {\mathbb{R}}$.

Under this transformation, one finds that
\begin{equation}
\lambda \rightarrow  {1 \over 2} (1+\cos 2 \rho) \lambda -{1 \over 2} (\cos 2 \rho -1)e^{2 i \theta}
\bar{\lambda} + {1 \over 2} \sin 2 \rho (w-{\bar{w}}) e^{i \theta}
\end{equation}
and one can always choose $\rho, \theta$ in order to make this expression vanish.

Having eliminated $B^{ij}$ there are a number of cases to consider.

\begin{itemize}

\item[(i)] Suppose ${\rm Im \ } w \neq 0$. Then one can set $\alpha=0$ and $\tau^i=0$
for all $i$,
by applying a ${\mathbb{R}}^9$ transformation generated by $R^i \Gamma_{+i} + R^{\bar{i}} \Gamma_{+ \bar{i}} + \xi \Gamma_{+ \sharp}$, where $\xi$ is fixed by
\begin{equation}
\sqrt{2} ({\rm Im \ } w) \bigg(1+{ {\bar{\psi}}^j \psi^j \over |w|^2}\bigg) \xi + {\rm Im \ } \bigg(\alpha - {1 \over w} \tau^j {\bar{\psi}}^j \bigg) =0~,
\end{equation}
and $R^i$ is then given by
\begin{equation}
R^j = {1 \over 2w} (\tau^j + \sqrt{2} \xi \psi^j )~.
\end{equation}
Note that this transformation in fact only sets ${\rm Im \ } \alpha =0$.
However, the real part of $\alpha$ can also be removed by subtracting a real multiple of
$\nu^1$ from $\nu^2$.

Then apply a $SU(4)$ transformation to set $\psi^2=\psi^3 = \psi^4 =0$ with $\psi^1=\psi \in{\mathbb{R}}$.
Next apply a $SU(3)$ transformation in the directions $2,3,4$ to eliminate the $e_{12}$ and $e_{13}$
terms, and set the $e_{14}$ coefficient to be real.
After applying all these transformations, one has
\begin{equation}
\nu^2 =x (e_5 + e_{12345})+i(e_5 - e_{12345}) + \psi (e_{15}+e_{2345}) + \mu (e_{14}-e_{23})~,
\end{equation}
where $x, \psi, \mu \in {\mathbb{R}}$.

\item[(ii)] Suppose ${\rm Im \ } w = 0$. Then $w e_5 + {\bar{w}} e_{12345}$ is $Spin(7)$ invariant
and by the reasoning given previously one can apply a $Spin(7)$ transformation to
set $A^{ij}=0$ for all $i,j$, whilst keeping $B^{ij}=0$ also.
To proceed there are then a number of sub-cases to consider.

\begin{itemize}

\item[(a)] If $\psi \neq 0$, then one can apply a ${\mathbb{R}}^8$ transformation,
with $R^i = \sigma \psi^i$ (and $\xi=0$) for appropriately chosen $\sigma \in {\mathbb{C}}$ in order
to set $\alpha=0$, whilst keeping $A^{ij}=0$.
Then apply a $SU(4)$ transformation to set $\psi^2=\psi^3=\psi^4=0$ and
take without loss of generality $\psi^1=1$. Then apply a $SU(3)$ transformation
in the $2,3,4$ directions to set $\tau^3=\tau^4=0$ with $\tau^2 \in {\mathbb{R}}$.
Finally, apply a ${\mathbb{R}}^9$ transformation with
\begin{equation}
R^1 =- {1 \over \sqrt{2}} \xi w, \quad R^2=R^3=R^4=0~,
\end{equation}
where
\begin{equation}
\xi =- {{\rm Re \ } \tau^1 \over \sqrt{2} (1+w^2)}~,
\end{equation}
which sets ${\rm Re \ } \tau^1 =0$.
The second normal then simplifies to
\begin{equation}
\nu^2 = y (e_5+e_{12345}) + e_{15}+e_{2345} + i \lambda (e_1 - e_{234})
+ \mu (e_2 - e_{134})~,
\end{equation}
for $y, \lambda, \mu \in {\mathbb{R}}$.

\item[(b)] If $\psi^i=0$ for all $i$ then there are two further possibilities.

In the first, $w \neq 0$, and one can use a ${\mathbb{R}}^8$ transformation to set $\tau^i=0$ for all $i$.
The second normal then simplifies to
\begin{equation}
\nu^2 = iy (1-e_{1234}) + e_5 + e_{12345}~,
\end{equation}
for $y \in {\mathbb{R}}$.

In the second, $w=0$. Then one can use a $SU(4)$ transformation to set $\tau^2=\tau^3=\tau^4=0$
with $\tau^1=\tau \in {\mathbb{R}}$, and the second normal spinor can be written as
\begin{equation}
\nu^2 = iy (1-e_{1234}) + \tau (e_1+e_{234})~.
\end{equation}

\end{itemize}

\end{itemize}

To summarize so far, the second normal can be written in
one of four possible canonical forms:

\begin{equation}
\label{anullx1}
\nu^2 =c_1 (e_5 + e_{12345})+i(e_5 - e_{12345}) + c_2 (e_{15}+e_{2345}) + c_3 (e_{14}-e_{23})~,
\end{equation}
or
\begin{equation}
\label{anullx2}
\nu^2 = k_1 (e_5+e_{12345}) + e_{15}+e_{2345} + i k_2 (e_1 - e_{234})
+ k_3 (e_2 - e_{134})~,
\end{equation}
or
\begin{equation}
\label{anullx3}
\nu^2 = ix (1-e_{1234}) + e_5 + e_{12345}~,
\end{equation}
or
\begin{equation}
\label{anullx4}
\nu^2 = iy (1-e_{1234}) + \tau (e_1+e_{234})~,
\end{equation}
where $c_1, c_2, c_3, k_1, k_2, k_3, x, y, \tau \in {\mathbb{R}}$.

Further simplification can be obtained by computing the norms of the vector field bilinears
associated with $\nu^2$ in the above four cases ({\ref{anullx1}})-({\ref{anullx4}}).
One finds $-16 c_3^2 (1+c_1^2)$, $-16(k_2^2+k_3^2)$, $-16x^2$ and $0$ respectively.
If any of these norms does not vanish, then the second normal is $SU(5)$ invariant. Since the two normals are un-ordered,
  the corresponding case has already
been considered   in the previous section. Therefore we demand that both normals are associated with null vectors and as a result
we set $c_3=0$, $k_2=k_3=0$ and $x=0$ in
({\ref{anullx1}})-({\ref{anullx3}}).

Using this, the cases ({\ref{anullx1}})-({\ref{anullx3}}) can be combined as
\begin{equation}
\label{anully1}
\nu^2 = b_1 (e_5 + e_{12345}) + i b_2 (e_5-e_{12345}) + b_3 (e_{15}+e_{2345})~,
\end{equation}
for $b_1, b_2, b_3 \in {\mathbb{R}}$. In fact, additional simplification to this case can be obtained by requiring
that the vector biliniear associated with $\nu^2+\nu^1$ be null. This forces $b_2=0$.

To summarize, when $\nu^1=1+e_{1234}$, and all possible real linear combinations of $\nu^1$ and
$\nu^2$ generate null vector fields, one can without loss of generality take
\begin{equation}
\label{anullt1}
\nu^2 = a(e_5 + e_{12345}) + b (e_{15}+e_{2345})~,
\end{equation}
or
\begin{equation}
\label{anullt2}
\nu^2 = i m (1-e_{1234}) + n (e_1+e_{234})~,
\end{equation}
for $a, b, m, n \in {\mathbb{R}}$.

\appendix{Analysis of $Spin(7)$ Solutions}

\setcounter{subsection}{0}

Before we proceed with the detailed analysis, we shall first introduce some notation. In particular, it will be convenient to define
\begin{eqnarray}
\label{extrad1}
(S_{N_1 N_2})_{M_1 M_2 M_3 M_4 M_5 M_6} &=& F_{N_1 [M_1 M_2 M_3} F_{|N_2| M_4 M_5 M_6]} \ ,
\nonumber \\
(Q_{N_1 N_2})_{M_1 M_2 M_3 M_4} &=& F_{L [ M_1 M_2 M_3} F_{M_4] N_1 N_2}{}^L \ .
\end{eqnarray}

It will also be useful to decompose the indices in a $2+9$ fashion. We use the null basis ${\bf{e}}^\pm = {1 \over \sqrt{2}}({\bf{e}}^5 \pm {\bf{e}}^0)$
and use the index notation ${\hat{N}}$ to denote any spacetime direction  apart from the lightcone $+$ and $-$. We also write

\begin{eqnarray}
\phi_{{\hat{N}}_1 {\hat{N}}_2 {\hat{N}}_3}&=& F_{+ {\hat{N}}_1 {\hat{N}}_2 {\hat{N}}_3}~,
\nonumber \\
\chi_{{\hat{N}}_1 {\hat{N}}_2 {\hat{N}}_3} &=& F_{- {\hat{N}}_1 {\hat{N}}_2 {\hat{N}}_3}~,
\nonumber \\
\omega_{{\hat{N}}_1 {\hat{N}}_2} &=& F_{+-{\hat{N}}_1 {\hat{N}}_2}~,
\nonumber \\
\psi_{{\hat{N}}_1 {\hat{N}}_2 {\hat{N}}_3 {\hat{N}}_4} &=& F_{{\hat{N}}_1 {\hat{N}}_2 {\hat{N}}_3 {\hat{N}}_4}~.
\end{eqnarray}

In all $Spin(7)$ cases, after a computer calculation, one finds that the tensors $S$ and $Q$ satisfy

\begin{eqnarray}
(Q_{+ {\hat{N}}_1})_{+ {\hat{N}}_2 {\hat{N}}_3 {\hat{N}}_4} &=&0 \ ,
\nonumber \\
(S_{+ {\hat{N}}})_{+ {\hat{N}}_1 {\hat{N}}_2 {\hat{N}}_3 {\hat{N}}_4 {\hat{N}}_5} &=&0 \ ,
\nonumber \\
(Q_{+-})_{+ {\hat{N}}_1 {\hat{N}}_2 {\hat{N}}_3}&=&0 \ ,
\nonumber \\
(Q_{{\hat{N}}_1 {\hat{N}}_2})_{+ {\hat{N}}_3 {\hat{N}}_4 {\hat{N}}_5} &=&0 \ ,
\nonumber \\
(S_{+ {\hat{N}}})_{{\hat{N}}_1 {\hat{N}}_2 {\hat{N}}_3 {\hat{N}}_4 {\hat{N}}_5 {\hat{N}}_6} &=&0 \ ,
\nonumber \\
(S_{+-})_{{\hat{N}}_1 {\hat{N}}_2 {\hat{N}}_3 {\hat{N}}_4 {\hat{N}}_5 {\hat{N}}_6} &=&0 \ .
\end{eqnarray}

To proceed, note that the constraint $(Q_{+ {\hat{N}}_1})_{+ {\hat{N}}_2 {\hat{N}}_3 {\hat{N}}_4}=0$ implies that
\begin{equation}
\phi_{\hat{L} [{\hat{N}}_2 {\hat{N}}_3} \phi_{{\hat{N}}_4] {\hat{N}}_1}{}^{\hat{L}}=0~.
\end{equation}
Hence $\phi$ are the structure constants of a Euclidean Lie algebra, $\mathfrak{g}$, of dimension 9.
The constraint $(S_{+ {\hat{N}}})_{+ {\hat{N}}_2 {\hat{N}}_3 {\hat{N}}_4 {\hat{N}}_5 {\hat{N}}_6}=0$ implies that
\begin{equation}
\label{auxcv1}
\phi_{[{\hat{N}}_2 {\hat{N}}_3 {\hat{N}}_4} \phi_{{\hat{N}}_5 {\hat{N}}_6] {\hat{N}}}=0~.
\end{equation}
Suppose that $\mathfrak{g}$ is not abelian. Then write $\mathfrak{g}=\mathfrak{g}_{ss}\oplus^{9-d} \mathfrak{u}(1)$,
where $\mathfrak{g}_{ss}$ is a semi-simple
Lie algebra of dimension $d$. Split the indices ${\hat{N}}$ as ${\hat{N}}= \{ i , \alpha \}$ where $i$ denote indices on $\mathfrak{g}_{ss}$,
and $\alpha$ are $\mathfrak{u}(1)$ indices. Then ({\ref{auxcv1}}) can be rewritten as
\begin{equation}
\label{auxcv2}
\phi_{\ell_2 \ell_3 \ell_4}\phi_{\ell_5 \ell_6 n}
- 3 \phi_{\ell_5 [\ell_2 \ell_3}\phi_{\ell_4] \ell_6 n}
+3  \phi_{\ell_6 [\ell_2 \ell_3}\phi_{\ell_4] \ell_5 n}
+3 \phi_{\ell_5 \ell_6 [\ell_2}\phi_{\ell_3 \ell_4] n} =0~.
\end{equation}

Suppose  $\mathfrak{g}_{ss}=su(2) \oplus su(2)$ or $\mathfrak{g}_{ss}=su(2) \oplus su(2) \oplus su(2)$, by taking
$\ell_2, \ell_3, \ell_4$ to lie in one $su(2)$, and $\ell_4, \ell_5, n$ to lie in another $su(2)$,
({\ref{auxcv2}}) implies that $\phi_{\ell_1 \ell_2 \ell_3}\phi_{\ell_5 \ell_6 n}=0$, which is a contradiction.
Next suppose that $\mathfrak{g}_{ss}=su(3)$, then by contracting ({\ref{auxcv2}}) with $\phi^{\ell_2 \ell_3 \ell_4}$,
one finds $\phi_{\ell_5 \ell_6 n}=0$, which again is a contradiction.
Hence, the only solution is  $\mathfrak{g}_{ss}=su(2)$ for which  ({\ref{auxcv1}}) holds
automatically. Therefore $\mathfrak{g}=\oplus^9\mathfrak{u}(1)$ or $\mathfrak{g}=\mathfrak{su}(2) \oplus^6 \mathfrak{u}(1)$.

To continue consider first the case $\mathfrak{g}=\mathfrak{su}(2) \oplus^6 \mathfrak{u}(1)$.
Examining various components of $T^4$ and $T^5$, we find
\begin{itemize}
\item[(i)] $(Q_{+-})_{+ \ell_1 \ell_2 \alpha}=0$ implies that $\omega_{i \alpha}=0$.
\item[(ii)] $(Q_{ij})_{+ \beta_1 \beta_2 \beta_3}=0$ implies that $\psi_{k \beta_1 \beta_2 \beta_3}=0$.
\item[(iii)] $(Q_{ij})_{+ \beta_1 \beta_2 \ell}=0$ implies that $\omega_{\beta_1 \beta_2}=0$~
and $\psi_{ij \beta_1 \beta_2}=0$.
\item[(iv)] $(S_{+ \beta})_{\ell_1 \ell_2 \ell_3 \beta_4 \beta_5 \beta_6}=0$ implies that $\psi_{\beta_1 \beta_2 \beta_3
\beta_4}=0$.
\item[(v)] $(S_{+-})_{\ell_1 \ell_2 \ell_3 \beta_1 \beta_2 \beta_3}=0$ implies that $\chi_{\beta_1 \beta_2 \beta_3}=0$.
\end{itemize}

Next, note that in the $Spin(7)$ case with $\nu^2=e_5+e_{12345}$, a computer calculation yields the additional
condition
\begin{equation}
\label{extrab1}
(S_{{\hat{N}}_1 {\hat{N}}_2})_{+- {\hat{N}}_3 {\hat{N}}_4 {\hat{N}}_5 {\hat{N}}_6}=0 \ .
\end{equation}
It is straightforward to show that the vanishing of $(S_{n_1 n_2})_{+- n_3 n_4 \alpha \beta}$  implies that $\chi_{n \alpha \beta}=0$.

For the remaining $Spin(7)$ cases described in section (4.1.2) one finds, after a computer calculation,
the additional condition
\begin{equation}
(Q_{+ {\hat{N}}})_{- {\hat{N}}_1 {\hat{N}}_2 {\hat{N}}_3} =0 \ .
\end{equation}
The vanishing of $(Q_{+i})_{- \alpha \beta j}$ again implies that $\chi_{n \alpha \beta}=0$.

To proceed further:

\begin{itemize}

\item[(a)] If $\nu^2=e_5+e_{12345}$, then as $\omega$ is a simple 2-form, one must have $(S_{+-})_{+- {\hat{N}}_1 {\hat{N}}_2 {\hat{N}}_3 {\hat{N}}_4}=0$. One evaluating this component of $S$, one finds that all $u$ vanish. Hence these solutions are locally maximally
supersymmetric.

\item[(b)] If $\nu^2=e_1+e_{234}$ or $\nu^2=i(1-e_{1234})+y(e_1+e_{234})$ (for $y \neq 0$), then a computer calculation
yields
\begin{equation}
(Q_{+-})_{- {\hat{N}}_1 {\hat{N}}_2 {\hat{N}}_3}=0
\end{equation}
which implies that
\begin{equation}
\chi_{i \beta [n_1} \omega_{n_2]}{}^i =0 \ .
\end{equation}
Suppose first that $\omega \neq 0$.
As $\omega$ is a simple 2-form, this implies that $\chi$ is a simple 3-form. Hence it follows that
\begin{equation}
(S_{- {\hat{N}}_1})_{- {\hat{N}}_2 {\hat{N}}_3 {\hat{N}}_3 {\hat{N}}_5 {\hat{N}}_6}=0 \ .
\end{equation}
It is straightforward to show that this implies that all $u=0$, and hence these solutions also are locally
maximally supersymmetric.

If, however $\omega=0$, then the vanishing of $(Q_{+i})_{- \alpha m n}$ implies that
$\chi_{n_1 n_2 \beta}=0$, and hence all components of $\phi$, $\psi$, $\omega$, $\chi$ are constrained to vanish with the exception of
$\phi_{\ell_1 \ell_2 \ell_3}$, $\psi_{\alpha \ell_1 \ell_2 \ell_3}$, $\chi_{\ell_1 \ell_2 \ell_3}$.
These conditions imply that the 4-form $F$ is simple, and hence $Q=0$ and $S=0$. However, $Q=0$ and $S=0$
are sufficient to force all remaining unfixed $u$ to vanish, hence these solutions are once more locally maximally supersymmetric.

\item[(c)] If $\nu^2=i(1-e_{1234})$, then again there are two subcases.
If $\omega=0$ then  the vanishing of $(Q_{+i})_{- \alpha m n}$ implies that
$\chi_{n_1 n_2 \beta}=0$, and hence all components of $\phi$, $\psi$, $\omega$, $\chi$ are constrained to vanish with the exception of
$\phi_{\ell_1 \ell_2 \ell_3}$, $\psi_{\alpha \ell_1 \ell_2 \ell_3}$, $\chi_{\ell_1 \ell_2 \ell_3}$.
These conditions imply that the 4-form $F$ is simple, and hence $Q=0$ and $S=0$. However, $Q=0$ and $S=0$
are sufficient to force all remaining unfixed $u$ to vanish, hence these solutions are  locally maximally supersymmetric.

If however, $\omega \neq 0$, then a computer calculation
yields
\begin{equation}
\label{jcx1}
(Q_{+-})_{- {\hat{N}}_1 {\hat{N}}_2 {\hat{N}}_3}=0
\end{equation}
which by the reasoning in $(b)$ again implies that
\begin{equation}
\label{gy1}
(S_{- {\hat{N}}_1})_{- {\hat{N}}_2 {\hat{N}}_3 {\hat{N}}_4 {\hat{N}}_5 {\hat{N}}_6}=0 \ .
\end{equation}
In addition, ({\ref{jcx1}}) implies that
\begin{equation}
\chi_{\alpha n_1 n_2} = V_\alpha \omega_{n_1 n_2}
\end{equation}
for some $V_\alpha$,
and note also that
\begin{equation}
\psi_{\alpha n_1 n_2 n_3}=W_\alpha \epsilon_{n_1 n_2 n_3} \ .
\end{equation}
Then the condition $(Q_{+i})_{- \alpha mn}=0$ implies that $V_\alpha$, $W_\alpha$ are linearly dependent.

It follows that the conditions on $\psi$ and $\chi$ obtained so far are sufficient to imply that
\begin{equation}
\label{gy2}
(S_{MN})_{{\hat{N}}_1 {\hat{N}}_2 {\hat{N}}_3 {\hat{N}}_4 {\hat{N}}_5 {\hat{N}}_6}=0 ,
\qquad (S_{{\hat{N}}_1 {\hat{N}}_2})_{\pm {\hat{N}}_3 {\hat{N}}_4 {\hat{N}}_5 {\hat{N}}_6 {\hat{N}}_7}=0 \ .
\end{equation}
On evaluating the conditions imposed on $u$ by ({\ref{gy1}}) and ({\ref{gy2}}), one finds
that all $u=0$, hence once again, the solutions are locally maximally supersymmetric.

\end{itemize}

The analysis of the case for which $\mathfrak{g}=\oplus^9\mathfrak{u}(1)$ (i.e. $\phi=0$) is more involved, and depends on
the various cases under consideration.

\subsection{Solutions with $\nu^2=e_5+e_{12345}$}

In order to analyse these solutions, note that the condition ({\ref{extrab1}}) implies (on contracting over the ${\hat{N}}_2, {\hat{N}}_3$ indices) that
\begin{equation}
\label{extrab2}
\omega_{\mu_1}{}^\lambda \psi_{\lambda \mu_2 \mu_3 \mu_4}  +3 \omega_{[\mu_2}{}^\lambda  \psi_{|\lambda \mu_1| \mu_3 \mu_4]} =0 \ .
\end{equation}
In addition, a computer calculation implies that
\begin{equation}
\label{extrab3}
(Q_{+-})_{\mu_1 \mu_2 \mu_3 \mu_4}=0
\end{equation}
which is equivalent to
\begin{equation}
\omega_{[\mu_1}{}^\lambda \psi_{|\lambda| \mu_2 \mu_3 \mu_4]} =0 \ .
\end{equation}
On comparing this equation with ({\ref{extrab2}}) one finds
\begin{equation}
\omega_{\mu_1}{}^\lambda \psi_{\lambda \mu_2 \mu_3 \mu_4}=0 \ .
\end{equation}
If $\omega \neq 0$, then this means there is a non-zero vector $v \in {\mathbb{R}}^9$ such that $i_v \psi=0$, and hence in particular
\begin{equation}
v^\lambda (S_{\lambda \mu_1})_{\mu_2 \mu_3 \mu_4 \mu_5 \mu_6 \mu_7} = v^\lambda (S_{\mu_1 \mu_2})_{\lambda \mu_3 \mu_4 \mu_5 \mu_6 \mu_7} =0 \ .
\end{equation}
By applying an $SU(4)$ gauge transformation, one can take without loss of generality $v^2=v^3=v^4=v^6=v^7=v^8=v^9=0$, then the above
condition forces the remaining degree of freedom in $u$ to vanish; such solutions are therefore locally maximally supersymmetric.

If, however, $\omega=0$, then this implies that
\begin{equation}
(S_{+ \lambda_1})_{- \lambda_2 \lambda_3 \lambda_4 \lambda_5 \lambda_6}=0 \ .
\end{equation}
On examining the components of this condition on the computer, one finds again that this condition
 forces the remaining degree of freedom in $u$ to vanish.

It follows that all solutions with $\nu^2=e_5+e_{12345}$ are locally maximally supersymmetric.

\subsection{Solutions with $\nu^2=im(1-e_{1234})+n(e_1+e_{234})$}

In order to analyse these solutions, note that a computer calculation yields the conditions
\begin{equation}
(S_{+-})_{+- \alpha_1 \alpha_2 \alpha_3 \alpha_4} =0, \qquad (Q_{+ \alpha_1})_{- \alpha_2 \alpha_3 \alpha_4}=0
\end{equation}
which imply that
\begin{equation}
\label{ccvt1}
\psi_{\lambda \alpha_2 \alpha_3 \alpha_4} \omega_{\alpha_1}{}^\lambda =0
\end{equation}
and
\begin{equation}
\omega  \wedge \omega =0 \ .
\end{equation}
This implies that $\omega$ is proportional to a simple 2-form on ${\mathbb{R}}^9$. We shall consider the cases
for which $\omega \neq 0$ and $\omega=0$ separately.

\subsubsection{Solutions with $\omega  \neq 0$}

To proceed, note that ({\ref{ccvt1}}) implies that there exists a non-vanishing vector field $v \in {\mathbb{R}}^9$ such that
\begin{equation}
i_v \Psi=0
\end{equation}
which in turn implies that
\begin{eqnarray}
\label{ccvt2}
v^\alpha (S_{\alpha \lambda_1})_{\lambda_2 \lambda_3 \lambda_4 \lambda_5 \lambda_6 \lambda_7}&=&0 \ ,
\nonumber \\
v^\alpha (S_{\lambda_1 \lambda_2})_{\alpha \lambda_3  \lambda_4 \lambda_5 \lambda_6 \lambda_7}&=&0 \ ,
\nonumber \\
v^\alpha (Q_{\alpha \lambda_1})_{\lambda_2 \lambda_3 \lambda_4 \lambda_5}&=&0 \ ,
\nonumber \\
v^\alpha (Q_{\lambda_1 \lambda_2})_{\alpha \lambda_3 \lambda_4 \lambda_5}&=&0 \ .
\end{eqnarray}

Consider first the case for which $\nu^2=i(1-e_{1234})$. In this case, one can use a $SU(4)$ transformation
to set $v^2=v^3=v^4=v^6=v^7=v^8=v^9=0$. A computer analysis of the conditions ({\ref{ccvt2}})
then implies sufficient conditions on the $u$ to impose the additional condition
\begin{equation}
(Q_{\lambda_1 \lambda_2})_{\lambda_3 \lambda_4 \lambda_5 \lambda_6}=0 \ .
\end{equation}
This constraint implies, using the result of \cite{3lienagy, 3liegp, 3liegut}, that one can write
\begin{equation}
\psi = k_1 \eta^1 + k_2 \eta^2~,
\end{equation}
where $\eta^1, \eta^2$ are two totally orthogonal simple 4-forms on ${\mathbb{R}}^9$.
The constraint $F \wedge F=0$ implies that $k_1 k_2=0$. Hence $\psi$ is proportional to a
simple 4-form on ${\mathbb{R}}^9$. It follows that
\begin{equation}
(S_{\lambda_1 \lambda_2})_{\lambda_3 \lambda_4 \lambda_5 \lambda_6 \lambda_7 \lambda_8}=0 \ .
\end{equation}
On evaulating the additional constraints on $u$ imposed by this condition, one finds, after a further computer calculation, that
\begin{equation}
\label{ccvt3}
(S_{+ \lambda_1})_{- \lambda_2 \lambda_3 \lambda_4 \lambda_5 \lambda_6}=0 \ .
\end{equation}
The conditions ({\ref{ccvt1}}) and ({\ref{ccvt3}}) are then sufficient to imply that
\begin{equation}
\psi=0 \ .
\end{equation}
In addition, from further computer calculation, one finds that
\begin{equation}
(Q_{-i})_{+- {\hat{\alpha}}_1 {\hat{\alpha}}_2}=0
\end{equation}
where $i,j$ correspond to the two directions associated with the simple 2-form $\omega$,
and ${\hat{\alpha}}_1, {\hat{\alpha}}_2$ are the orthogonal directions.
This implies that $\chi_{j {\hat{\alpha}}_1 {\hat{\alpha}}_2}=0$.
Furthermore, $\psi=0$ implies that
\begin{eqnarray}
(S_{M \alpha_1})_{\alpha_2 \alpha_3 \alpha_4 \alpha_5 \alpha_6 \alpha_7}&=&0
\nonumber \\
(S_{\alpha_1 \alpha_2})_{M \alpha_3 \alpha_4 \alpha_5 \alpha_6 \alpha_7}&=&0
\end{eqnarray}
for all $M$. On evaluating the extra constraints on $u$ obtained from these conditions, one finds
sufficient conditions to imply that
\begin{equation}
(Q_{-i})_{j {\hat{\alpha}}_1 {\hat{\alpha}}_2 {\hat{\alpha}}_3}=0
\end{equation}
which in turn implies that $\chi_{{\hat{\alpha}}_1 {\hat{\alpha}}_2 {\hat{\alpha}}_3}=0$.  So, all components of $\chi$ must vanish, with the exception of
$\chi_{ij {\hat{\alpha}}}$. This implies that $F$ is simple, and hence $Q=0$ and $S=0$. These solutions are therefore
locally maximally supersymmetric.

For the remaining $Spin(7)$ cases (with $\nu^2=im(1-e_{1234})+n(e_1+e_{234})$) a more straightforward computer calculation
yields directly the following constraints
\begin{eqnarray}
(S_{+ \lambda_1})_{- \lambda_2 \lambda_3 \lambda_4 \lambda_5 \lambda_6}&=&0 \ ,
\nonumber \\
(Q_{\lambda_1 \lambda_2})_{\lambda_3 \lambda_4 \lambda_5 \lambda_6} &=&0 \ ,
\nonumber \\
(Q_{- i})_{+- {\hat{\alpha}}_1 {\hat{\alpha}}_2} &=&0 \ ,
\nonumber \\
(S_{+-})_{- ij {\hat{\alpha}}_1 {\hat{\alpha}}_2 {\hat{\alpha}}_3} &=&0 \ .
\end{eqnarray}
As in the previous analysis, the first two of these conditions imply that $\psi=0$, whereas the last two conditions imply
that all components of $\chi$ must vanish, with the exception of
$\chi_{ij {\hat{\alpha}}}$. This implies that $F$ is simple, and hence $Q=0$ and $S=0$. These solutions are therefore again
locally maximally supersymmetric.

\subsubsection{Solutions with $\omega = 0$}

To proceed, we first consider the cases for which $\nu^2=e_1+e_{234}$ or $\nu^2=i(1-e_{1234})+y(e_1+e_{234})$
for $y \in {\mathbb{R}}$, $y \neq 0$. Note that a computer calculation yields the condition
\begin{equation}
(Q_{\lambda_1 \lambda_2})_{\lambda_3 \lambda_4 \lambda_5 \lambda_6} =0 \ .
\end{equation}

If $\psi \neq 0$, then this condition, together with $F \wedge F=0$, implies that $\psi$ is a simple 4-form on ${\mathbb{R}}^9$.
We therefore split the indices in a $4+5$ fashion as $\lambda = \{i , {\hat{\alpha}} \}$, where $i$ denote the 4 indices in
the directions of $\psi$, and ${\hat{\alpha}}$ denote the remaining 5 directions.
Note that $F \wedge F=0$ implies that
\begin{equation}
\chi_{{\hat{\alpha}}_1 {\hat{\alpha}}_2 {\hat{\alpha}}_3} =0 \ .
\end{equation}
Furthermore, a computer calculation yields the condition
\begin{equation}
(T^4_{L N_1})_{N_2 N_3 N_4}{}^L=0
\end{equation}
from which one finds
\begin{equation}
F_{L_1 L_2 [ij}F^{L_1 L_2}{}_{{\hat{\alpha}}] -}=0
\end{equation}
which implies
\begin{equation}
\chi_{mn {\hat{\alpha}}}=0 \ .
\end{equation}
A computer calculation also implies that
\begin{equation}
(S_{-m})_{{\hat{\alpha}}_1 {\hat{\alpha}}_2 i \ell_1 \ell_2 \ell_3} =0
\end{equation}
which in turn implies that
\begin{equation}
\chi_{i {\hat{\alpha}}_1 {\hat{\alpha}}_2}=0 \ .
\end{equation}
It follows that the only nonzero components of $\chi$ are $\chi_{\ell_1 \ell_2 \ell_3}$, and therefore $F$ is simple.
Therefore, for these solutions $Q=0$ and $S=0$, which implies that they are locally maximally supersymmetric.

It remains to consider the case when $\psi=0$. Then the only non-zero components of $S$ are
$(S_{- \alpha_1})_{- \alpha_2 \alpha_3 \alpha_4 \alpha_5 \alpha_6}$. There are a number of subcases to consider,

Firstly, if $(S_{- \alpha})_{- \alpha_1 \alpha_2 \alpha_3 \alpha_4}{}^\alpha \neq 0$, then
the constraint
\begin{equation}
(S_{- [\beta_1|})_{- [\alpha_1 \alpha_2 \alpha_3 \alpha_4 \alpha_5} \chi_{\alpha_6 \alpha_7] |\beta_2]} =0
\end{equation}
is sufficient to imply that either all $u$ vanish, or $\chi=0$. In both cases, this implies the solutions are locally maximally supersymmetric. Secondly, if $(S_{- \alpha})_{- \alpha_1 \alpha_2 \alpha_3 \alpha_4}{}^\alpha = 0$
then this condition reduced the number of degrees of freedom in the $u$ from 3 to 2, and implies that
$\chi_{\alpha_1 \alpha_2 \alpha_3}$ are the structure constants of a 9-dimensional Euclidean Lie algebra
${\mathfrak{h}}$. If ${\mathfrak{h}}$ is not semi-simple then there exists nonzero $v \in {\mathbb{R}}^9$ such that
\begin{equation}
\label{vcb1}
v^\alpha (S_{- \alpha})_{- \beta_1 \beta_2 \beta_3 \beta_4 \beta_5}=0
\end{equation}
and
\begin{equation}
\label{vcb2}
v^\alpha (S_{- \beta_1})_{- \alpha \beta_2 \beta_3 \beta_4 \beta_5}=0 \ .
\end{equation}
By making an appropriately chosen $SU(3)$ transformation which leaves $\nu^1, \nu^2$ invariant, one can take,
without loss of generality, $v^3=v^4=v^7=v^8=v^9=0$. Then ({\ref{vcb1}}) and ({\ref{vcb2}}) imply that all the $u$
vanish, so the solutions are locally maximally supersymmetric.
If, however, $\mathfrak{h}$ is semi-simple, one must have ${\mathfrak{h}}= {\mathfrak{su}}(2) \oplus {\mathfrak{su}}(2)
\oplus {\mathfrak{su}}(2)$; but there exists a nonzero $v \in {\mathbb{R}}$ such that ({\ref{vcb1}}) holds, which is not possible in
the case ${\mathfrak{h}}= {\mathfrak{su}}(2) \oplus {\mathfrak{su}}(2)
\oplus {\mathfrak{su}}(2)$. It follows that ${\mathfrak{h}}$ cannot be semi-simple.

Hence, we have shown that if  $\nu^2=e_1+e_{234}$ or $\nu^2=i(1-e_{1234})+y(e_1+e_{234})$, the solutions must
all be locally maximally supersymmetric. It remains to consider the solutions with $\nu^2=i(1-e_{1234})$. For these
solutions, observe that $\phi=0$ and $\omega=0$ implies that:
\begin{eqnarray}
(S_{+ {\hat{N}}_1})_{- {\hat{N}}_2 {\hat{N}}_3 {\hat{N}}_4 {\hat{N}}_5 {\hat{N}}_6} &=&0 \ ,
\nonumber \\
(S_{+-})_{N_1 N_2 N_3 N_4 N_5 N_6} &=&0 \ ,
\nonumber \\
(S_{N_1 N_2})_{+- N_3 N_4 N_5 N_6} &=&0 \ .
\end{eqnarray}
A computer computation shows that these conditions are sufficient to reduce the 78 degrees of freedom in $u$ to 30.
Furthermore, one obtains the conditions
\begin{equation}
(Q_{{\hat{N}}_1  {\hat{N}}_2})_{{\hat{N}}_3 {\hat{N}}_4 {\hat{N}}_5 {\hat{N}}_6} =0
\end{equation}
and
\begin{equation}
(S_{- {\hat{N}}_1})_{{\hat{N}}_2 {\hat{N}}_3 {\hat{N}}_4 {\hat{N}}_5 {\hat{N}}_6 {\hat{N}}_7} =0 \ .
\end{equation}
Then, from the reasoning used to analyse the solutions with $\nu^2=e_1+e_{234}$ or $\nu^2=i(1-e_{1234})+y(e_1+e_{234})$, one finds that if $\psi \neq 0$ then the solutions are locally maximally supersymmetric. Therefore, consider the remaining case, with $\psi=0$. For such solutions, one must also have
\begin{equation}
(S_{{\hat{N}}_1 {\hat{N}}_2})_{- {\hat{N}}_3 {\hat{N}}_4 {\hat{N}}_5 {\hat{N}}_6 {\hat{N}}_7}=0
\end{equation}
and these conditions are sufficient to reduce the numbers of degrees of freedom in $u$ further, from 30 to 18.
It will be convenient to split the indices in an $8+1$ fashion as ${\hat{N}} = \{ i , \sharp \}$,
where $i,j=1,2,3,4,6,7,8,9$, and let $\alpha, {\bar{\alpha}}$ denote $SU(4)$ holomorphic and antiholomorphic
indices in these 8 directions. A computer computation implies that the only non-vanishing
component of $(S_{-i})_{- j_1 j_2 j_3 j_4 j_5}$ is, up to complex conjugation,
$(S_{- \alpha})_{- \mu_1 \mu_2 \mu_3 \mu_4 {\bar{\beta}}}$, and moreover
\begin{equation}
\label{vcb3}
\chi_{\alpha [\mu_1 \mu_2} \chi_{\mu_3 \mu_4 {\bar{\beta}}]} = -{9 \over 10} \xi \delta_{\alpha \bar{\beta}}
\epsilon_{\mu_1 \mu_2 \mu_3 \mu_4}
\end{equation}
where $\xi$ is linear in $u$. However, note that one can use a $SU(4)$ transformation, which leaves
$\nu^1$, $\nu^2$ invariant to set $\chi_{124}=\chi_{134}=\chi_{234}=0$ (in holomorphic indices).
It is then straightforward to show that ({\ref{vcb3}}) implies that $\xi=0$. This imposes additional conditions on
$u$ and reduces further the number of degrees of freedom from 18 to 16. Furthermore, one finds
\begin{equation}
\label{vcb4}
\chi_{i [j_1 j_2} \chi_{j_3 j_4 j_5]} =0 \ .
\end{equation}
Note that ({\ref{vcb4}}) implies that $\chi_{ijk}$ are the structure constants of an 8-dimensional Euclidean Lie
algebra ${\mathfrak{h}}$. As ({\ref{vcb4}}) does not hold for ${\mathfrak{h}}= {\mathfrak{su}}(3)$
or ${\mathfrak{h}} = {\mathfrak{su}}(2) \oplus {\mathfrak{su}}(2) \oplus^2 {\mathfrak{u}}(1)$, the remaining possibilities
are ${\mathfrak{h}} = {\mathfrak{su}}(2) \oplus^5 {\mathfrak{u}}(1)$ or ${\mathfrak{h}} = \oplus^8 {\mathfrak{u}}(1)$.
Also observe that a computer computation can be used to show that all of the previous constraints imposed on $u$ are sufficient to imply
\begin{equation}
(S_{- \sharp})_{- i_1 i_2 i_3 i_4 \sharp}=0
\end{equation}
which implies that $\chi_{\sharp ij}$ defines a simple 2-form on ${\mathbb{R}}^8$. Hence, there exists nonzero $v \in {\mathbb{R}}^8$ such that
\begin{equation}
v^i \chi_{ijk}=0, \qquad v^i \chi_{\sharp ij} =0
\end{equation}
which in turn implies
\begin{eqnarray}
\label{vcb5}
v^i (S_{- \sharp})_{- i j_1 j_2 j_3 j_4} &=&0 \ ,
\nonumber \\
v^i (S_{- j_1})_{- \sharp i j_2 j_3 j_4} &=&0 \ ,
\nonumber \\
v^i (S_{-i})_{- \sharp j_1 j_2 j_3 j_4} &=&0 \ .
\end{eqnarray}
By applying a $SU(4)$ transformation which leaves $\nu^1, \nu^2$ invariant, one can take, without loss of
generality $v^2=v^3=v^4=v^6=v^7=v^8=v^9=0$, then it is straightforward to show using a further computer calculation,
that ({\ref{vcb5}}) is sufficient to imply that all $u$ vanish. Such solutions are therefore also locally maximally supersymmetric.

\appendix{Analysis of $SU(5)$ solutions with $\nu^2=i(1-e_{12345})$}

To analyse these solutions, it is convenient to split the indices in a $10+1$ fashion and write
$N= \{ 0, {\hat{N}} \}$ where ${\hat{N}} \neq 0$. Also, define
\begin{eqnarray}
\phi_{{\hat{N}}_1 {\hat{N}}_2 {\hat{N}}_3} &=& F_{0 {\hat{N}}_1 {\hat{N}}_2 {\hat{N}}_3} \ ,
\nonumber \\
\psi_{{\hat{N}}_1 {\hat{N}}_2 {\hat{N}}_3 {\hat{N}}_4} &=& F_{{\hat{N}}_1 {\hat{N}}_2 {\hat{N}}_3 {\hat{N}}_4}
\end{eqnarray}
and $Q$ and $S$ are also defined as in ({\ref{extrad1}}).
A computer calculation yields the following conditions on $Q$ and $S$;
\begin{equation}
\label{su5x1}
(Q_{0 {\hat{N}}_1})_{0 {\hat{N}}_2 {\hat{N}}_3 {\hat{N}}_4} =0
\end{equation}
and
\begin{equation}
\label{su5x2}
(S_{0 [{\hat{N}}_1})_{|0| {\hat{N}}_2] {\hat{N}}_3 {\hat{N}}_4 {\hat{N}}_5 {\hat{N}}_6} + (Q_{{\hat{N}}_1 {\hat{N}}_2})_{{\hat{N}}_3 {\hat{N}}_4 {\hat{N}}_5 {\hat{N}}_6} =0 \ .
\end{equation}
Note that ({\ref{su5x1}}) implies that $\phi_{{\hat{N}}_1 {\hat{N}}_2 {\hat{N}}_3}$ are the structure constants of a 10-dimensional
Euclidean Lie algebra ${\mathfrak{g}}$, whereas ({\ref{su5x2}}) can be rewritten as
\begin{equation}
\label{su5x3}
\psi_{{\hat{L}} [{\hat{N}}_3 {\hat{N}}_4 {\hat{N}}_5} \psi_{{\hat{N}}_6] {\hat{N}}_1 {\hat{N}}_2}{}^{{\hat{L}}}
-{4 \over 5} \phi_{{\hat{N}}_1 {\hat{N}}_2 [{\hat{N}}_3} \phi_{{\hat{N}}_4 {\hat{N}}_5 {\hat{N}}_6]} =0 \ .
\end{equation}

There are two cases to consider, according as ${\mathfrak{g}}$ is semi-simple or not semi-simple.

\begin{itemize}

\item[(i)] Suppose ${\mathfrak{g}}$ is not semi-simple. Then there exists nonzero $v \in {\mathbb{R}}^{10}$ such that
$i_v \phi=0$, and this, together with $F \wedge F=0$, implies that
\begin{eqnarray}
\label{su5x4}
v^{\hat{N}} (S_{0 {\hat{N}}})_{0 {\hat{N}}_1 {\hat{N}}_2 {\hat{N}}_3 {\hat{N}}_4 {\hat{N}}_5} &=&0 \ ,
\nonumber \\
v^{\hat{N}} (S_{0 {\hat{N}}_1})_{0 {\hat{N}} {\hat{N}}_2 {\hat{N}}_3 {\hat{N}}_4 {\hat{N}}_5} &=&0 \ ,
\nonumber \\
v^{\hat{N}} (Q_{0 {\hat{N}}})_{{\hat{N}}_1 {\hat{N}}_2 {\hat{N}}_3 {\hat{N}}_4} &=&0 \ ,
\nonumber \\
v^{\hat{N}} (S_{0 {\hat{N}}})_{{\hat{N}}_1 {\hat{N}}_2 {\hat{N}}_3 {\hat{N}}_4 {\hat{N}}_5 {\hat{N}}_6} &=&0 \ .
\end{eqnarray}
Without loss of generality, one can make a $SU(5)$ gauge transformation, which leaves $\nu^1, \nu^2$ invariant,
to set $v^j=0$ for $j \neq 1$. After some computer analysis, one finds that the resulting conditions on $u$ are sufficient to
imply that
\begin{equation}
(S_{0 {\hat{N}}_1})_{0 {\hat{N}}_2 {\hat{N}}_3 {\hat{N}}_4 {\hat{N}}_5 {\hat{N}}_6}=0 \ .
\end{equation}
On substituting this condition into ({\ref{su5x3}}) one finds
\begin{equation}
\label{su5x5}
\phi_{{\hat{N}}_1 {\hat{N}}_2 [{\hat{N}}_3} \phi_{{\hat{N}}_4 {\hat{N}}_5 {\hat{N}}_6]} =0
\end{equation}
and
\begin{equation}
\phi_{{\hat{N}}_1 [{\hat{N}}_3 {\hat{N}}_4} \phi_{{\hat{N}}_5 {\hat{N}}_6] {\hat{N}}_2}=0
\end{equation}
and
\begin{equation}
\label{su5x6}
\psi_{{\hat{L}} [{\hat{N}}_3 {\hat{N}}_4 {\hat{N}}_5} \psi_{{\hat{N}}_6] {\hat{N}}_1 {\hat{N}}_2}{}^{{\hat{L}}} =0 \ .
\end{equation}

In particular, ({\ref{su5x6}}) implies, together with $F \wedge F=0$, that $\psi$ is a simple 1-form on ${\mathbb{R}}^{10}$;
and ({\ref{su5x5}}) implies that $\phi$ is proportional to a simple 3-form on ${\mathbb{R}}^{10}$. There are therefore two possibilities.
In the first, $\phi=0$ and ${\mathfrak{g}}= \oplus^{10} {\mathfrak{u}}(1)$; then $F$ is simple and $Q=0, S=0$. Such solutions are locally maximally supersymmetric. In the second, ${\mathfrak{g}} = {\mathfrak{su}}(2) \oplus^{7} {\mathfrak{u}}(1)$.
For this case, there must exits nonzero $v \in {\mathbb{R}}^{10}$ such that
\begin{eqnarray}
v^{\hat{N}} (S_{{\hat{N}} L_1})_{L_2 L_3 L_4 L_5 L_6 L_7} &=&0 \ ,
\nonumber \\
v^{\hat{N}} (S_{L_1 L_2})_{{\hat{N}} L_3 L_4 L_5 L_6 L_7}&=&0 \ ,
\nonumber \\
v^{\hat{N}} (Q_{{\hat{N}} L_1})_{L_2 L_3 L_4 L_5} &=&0 \ ,
\nonumber \\
v^{\hat{N}} (Q_{L_1 L_2})_{{\hat{N}} L_3 L_4 L_5} &=&0 \ .
\end{eqnarray}
These conditions are sufficient to imply that all $u$ vanish, hence these solutions are also locally
maximally supersymmetric.

\item[(ii)] Suppose ${\mathfrak{g}}$ is semi-simple, i.e. ${\mathfrak{g}}= {\mathfrak{so}}(5)$. Let $\alpha, {\bar{\beta}}$
denote holomorphic/antiholomorphic $SU(5)$ indices. A computer calculation yields the condition
\begin{equation}
(S_{0 [\alpha})_{|0| {\bar{\beta}}] {\hat{N}}_1 {\hat{N}}_2 {\hat{N}}_3 {\hat{N}}_4}=0
\end{equation}
which implies that
\begin{equation}
\phi_{\alpha {\bar{\beta}}  [{\hat{N}}_1} \phi_{{\hat{N}}_2 {\hat{N}}_3 {\hat{N}}_4]} =0 \ .
\end{equation}
On contracting this expression with $\phi^{{\hat{N}}_1 {\hat{N}}_2 {\hat{N}}_3}$, one finds that
\begin{equation}
\phi_{\alpha {\bar{\beta}} {\hat{N}}}=0
\end{equation}
i.e. $\phi$ is a $(3,0)+(0,3)$ form. Using the reasoning set out in the Appendix of
\cite{mtheor1}, one can make a $SU(5)$ gauge transformation which leaves $\nu^1, \nu^2$ invariant, and
take
\begin{equation}
\phi = \lambda_1 ({\bf{e}}^{125}+{\bf{e}}^{\bar{1} \bar{2} \bar{5}})
+ \lambda_2 ({\bf{e}}^{345}+{\bf{e}}^{\bar{3} \bar{4} \bar{5}})
\end{equation}
for $\lambda_1, \lambda_2 \in {\mathbb{R}}$. However, this does  not satisfy the Jacobi identity unless $\lambda_1=\lambda_2=0$,
in contradiction with the original assumption that $\phi$ are the structure constants of $ {\mathfrak{so}}(5)$.
Hence, there are no solutions for which ${\mathfrak{g}}$ is semi-simple.

\end{itemize}

\appendix{Plane wave solutions with $(SO(3)\times SU(3)\times U(1))\ltimes H_9$ symmetry}

We begin the analysis with a more general ansatz than that of equation (\ref{pw28}). In particular consider
\begin{equation}
ds^2 = 2 e^- e^+ + ds^2({\mathbb{R}}^9)~,~~~F = e^- \wedge \Phi~,~~~
\end{equation}
where
\begin{equation}
e^- = dv, \qquad e^+ = du + H dv,\qquad ds^2({\mathbb{R}}^9)=\delta_{ab} e^a e^b,\qquad e^a = dx^a~,~~~
\end{equation}
$H$ is a function
only of $x^a$, $a=1,6, \sharp, 2,3,4,7,8,9$,  $H=H(x)$, and $\Phi$ is a {\it constant} 3-form on ${\mathbb{R}}^9$.

To investigate the Killing spinor equations first observe that the only non-vanishing component of the spin connection is
\begin{equation}
\Omega_{-,-a} = {\partial H \over \partial x^a}~.
\end{equation}
Next the $+$ component of the KSE implies that $\partial_u\epsilon=0$ and the $a$ components can be solved to yield
\begin{equation}
\label{sol1x}
\epsilon = \bigg( 1 - \Gamma_+  x^a \big({1 \over 74} \Gamma_a{}^{b_1 b_2  b_3}  \Phi_{b_1 b_2 b_3}
-{1 \over 12} \Phi_{a b_1 b_2} \Gamma^{b_1 b_2} \big) \bigg) \eta
\end{equation}
where $\eta$ depends only on $v$. Assuming that $H$ is quadratic in the Euclidean coordinates $x$, $H(x)={1\over2} \lambda_{ab} x^a x^b$,
 one finds that the the $x$-independent terms in the $-$ component of the Killing spinor equations give
\begin{eqnarray}
\label{st2}
{d \eta_+ \over  d v}  +{1 \over 24}  \Gamma^{b_1 b_2 b_3} \Phi_{b_1 b_2 b_3}
\eta_+ &=& 0
\nonumber \\
{d \eta_- \over  d v}  +{1 \over 72}  \Gamma^{b_1 b_2 b_3} \Phi_{b_1 b_2 b_3}
\eta_- &=& 0
\end{eqnarray}
where $\eta = \eta_+ + \eta_-$ and $\Gamma_{\pm} \eta_{\pm}=0$,
while the $x$-dependent terms give the algebraic equation
\begin{equation}
\label{alg1}
V_a \Gamma_+ \eta_-=0~,
\end{equation}
where
\begin{eqnarray}
V_a &=& {1 \over 2592} \Gamma_a \bigg( \Gamma^{b_1 b_2 b_3} \Phi_{b_1 b_2 b_3} \bigg)^2
-{1 \over 576} \Phi_{a b_1 b_2} \Gamma^{b_1 b_2} \Gamma^{c_1 c_2 c_3} \Phi_{c_1 c_2 c_3}
\nonumber \\
&+&{1 \over 192}  \Gamma^{c_1 c_2 c_3} \Phi_{c_1 c_2 c_3}\Phi_{a b_1 b_2} \Gamma^{b_1 b_2}
- {1\over2}\lambda_{ab} \Gamma^b~.
\end{eqnarray}
The equations (\ref{st2}) are first order and always have solutions for any choice of $\Phi$. In particular there are at least 16 Killing spinors
given by the solutions for $\eta_+$. There may be additional Killing spinors provided $V_a$ has a non-trivial kernel. For the plane wave
solution to preserve 28 supersymmetries the kernel of $V_a$ must be the same for all $a$ and have dimension 12.

To continue let us specialize to the ansatz given in (\ref{pw28}). To be more specific, we introduce the hermitian basis in ${\mathbb{C}}^6={\mathbb{R}}^6\otimes {\mathbb{C}}$
as
\begin{equation}
{\bf{e}}^\alpha = {1 \over \sqrt{2}} (dx^{\alpha}+idx^{\alpha+5}),~~~\alpha=2,3,4
\end{equation}
and ${\bf{e}}^{\bar{\alpha}}$ is defined as the complex conjugate of ${\bf{e}}^\alpha$. In this basis, $\Phi$ and $(\lambda_{ab})$ in (\ref{pw28})
can be written as
\begin{eqnarray}
\Phi=k\, e^1\wedge e^6\wedge e^\sharp+ \mu\, {\bf{e}}^2\wedge {\bf{e}}^3\wedge {\bf{e}}^4+ \bar \mu\, {\bf{e}}^{\bar 2}\wedge {\bf{e}}^{\bar 3}\wedge {\bf{e}}^{\bar 4}~,
\end{eqnarray}
and
\begin{equation}
\lambda_{ij} = \lambda_1 \delta_{ij}, \qquad \lambda_{\alpha \bar{\beta}}= \lambda_2 \delta_{\alpha \bar{\beta}}
\end{equation}
for  $i,j=1,6,\sharp$, respectively, where   $k, \lambda_1,  \lambda_2$ are constant real parameters and $\mu$ is complex.
Observe  $\Phi$ and $(\lambda_{ab})$ are the most general 4-form and quadratic form, respectively, invariant under the $SO(3)\times SU(3)$
of the plane wave, see also \cite{hg, michelson, pope}.

To proceed, consider  $V_i \zeta=0$, where $\zeta=\Gamma_-\eta$. Taking $\Gamma_{(i)} V_{(i)}\zeta=0$,
where there is no summation over the indices in the parenthesis, one finds that it can be expressed as

\begin{equation}
\label{bt2}
\bigg( -{1 \over 18} k^2 - {\lambda_1\over2} -{1 \over 36}|\mu|^2 (1+ \Gamma_{2\bar2} \Gamma_{3\bar3} + \Gamma_{2\bar2} \Gamma_{4\bar4}
+\Gamma_{3\bar3} \Gamma_{4\bar4}) +{1 \over 24}k \Gamma_{16\sharp} (\mu\, \Gamma_{\bar2\bar3\bar4}+ \bar{\mu}\, \Gamma_{234})  \bigg) \zeta=0 \ .
\end{equation}
Moreover,
\begin{eqnarray}
\label{hol1}
V_\alpha\zeta&=&\bigg[ {1 \over 72} \Gamma_\alpha \big(-k^2-2|\mu|^2  (1+ \Gamma_{2\bar2} \Gamma_{3\bar3} + \Gamma_{2\bar2} \Gamma_{4\bar4}
+\Gamma_{3\bar3} \Gamma_{4\bar4})  \big) +{1 \over 48} \mu k \Gamma_{16\sharp} \epsilon_{\alpha \beta_1 \beta_2} \Gamma^{\beta_1 \beta_2}
\cr
&-&{1 \over 96}|\mu|^2 \epsilon_{\alpha \beta_1 \beta_2} \Gamma^{\beta_1 \beta_2} \Gamma_{234}
+{1 \over 32} |\mu|^2 \Gamma_{234} \epsilon_{\alpha \beta_1 \beta_2} \Gamma^{\beta_1 \beta_2} - {\lambda_2\over2} \Gamma_\alpha \bigg] \zeta=0
\end{eqnarray}
and  $V_{\bar{\alpha}} \zeta={\bar {(V_{\alpha})}}\zeta=0$.

To proceed, consider the cases.

\begin{enumerate}

\item[(i)] Suppose $\mu, k \neq 0$, then
\begin{equation}
\Gamma_{(\alpha)} V_{(\alpha)} \zeta=0, \qquad \Gamma_{(\bar{\alpha})} V_{(\bar{\alpha})} \zeta=0
\end{equation}
imply that
\begin{equation}
\epsilon_{(\alpha) \beta_1 \beta_2} \Gamma_{(\alpha)} \Gamma^{\beta_1 \beta_2} \zeta=0, \qquad \epsilon_{(\bar{\alpha}) \bar{\beta}_1 \bar{\beta}_2}
\Gamma_{(\bar{\alpha})} \Gamma^{\bar{\beta}_1 \bar{\beta}_2} \zeta =0~,
\end{equation}
which in turn give
\begin{equation}
\label{chop1}
\Gamma_{2\bar2} \zeta = \Gamma_{3\bar3} \zeta = \Gamma_{4\bar4} \zeta~.
\end{equation}
These give 2 independent and commuting  conditions on $\zeta$ each breaking half of the supersymmetry. This in particular implies that
the kernel of $V_a$ has dimension of at most 4. Thus such backgrounds cannot preserve 28 supersymmetries.

\item[(ii)] Suppose that $\mu=0$. Then it is straightforward to show that ({\ref{alg1}}) is equivalent to
\begin{equation}
\big(-{1 \over 9}k^2 - \lambda_1 \big) \zeta=0, \qquad \big(-{1 \over 36}k^2 - \lambda_2 \big) \Gamma_\alpha \zeta=0,
\qquad \big(-{1 \over 36}k^2 - \lambda_2 \big) \Gamma_{\bar{\alpha}} \zeta=0 \ .
\end{equation}
Thus either the kernel of $V_a$ is trivial and so the plane wave preserves 16 supersymmetries or
 $\lambda_1 = -{1 \over 9}k^2$ and $\lambda_2 = -{1 \over 36} k^2$ in which case the kernel of $V_a$ is 16-dimensional
 and the background is the maximally supersymmetric plane wave \cite{kg}.

\item[(iii)] Suppose that $k=0$. Then ({\ref{alg1}}) can be rewritten as

\begin{equation}
\big({1 \over 36}|\mu|^2  (1+  \Gamma_{2\bar2} \Gamma_{3\bar3} + \Gamma_{2\bar2} \Gamma_{4\bar4}
+\Gamma_{3\bar3} \Gamma_{4\bar4})  +{1\over2} \lambda_1 \big) \zeta=0~,
\end{equation}
\begin{equation}
\label{ev1}
V_\alpha\zeta=\big({1\over2} (\lambda_1 - \lambda_2) \Gamma_\alpha -{|\mu|^2 \over 96} \epsilon_{\alpha \beta_1 \beta_2} \Gamma^{\beta_1 \beta_2}
\Gamma_{234} +{|\mu|^2 \over 32} \Gamma_{234} \epsilon_{\alpha \beta_1 \beta_2} \Gamma^{\beta_1 \beta_2} \big) \zeta =0~,
\end{equation}
and $V_{\bar\alpha}\zeta=\bar{(V_\alpha)}\zeta=0$.

Observe that if $\mu=0$, then either the solution preserves 16 supersymmetries or  $\lambda_1=\lambda_2=0$ and so it is Minkowski space.
Next assuming that $\mu\not=0$, we consider $\Gamma_{(\bar\alpha)} V_{(\alpha)}\zeta+\Gamma_{(\alpha)} V_{(\bar\alpha)}\zeta=0$ to find
\begin{eqnarray}
\big( \lambda_1-\lambda_2 -{1 \over 12} |\mu|^2 (\Gamma_{\bar2\bar3\bar4} \Gamma_{234}+ \Gamma_{234} \Gamma_{\bar2\bar3\bar4})
\nonumber  \\
+{1 \over 32}|\mu|^2 \epsilon_{(\bar{\alpha}) \bar{\delta}_1 \bar{\delta}_2}
\epsilon_{(\alpha) \beta_1 \beta_2} (\Gamma^{\bar{\delta}_1 \bar{\delta}_2} \Gamma^{\beta_1 \beta_2}
+ \Gamma^{\beta_1 \beta_2}\Gamma^{\bar{\delta}_1 \bar{\delta}_2}) \big) \zeta=0~.
\end{eqnarray}
It is straightforward to see that these conditions imply again ({\ref{chop1}}) and so the kernel of $V_a$ has dimension at most 4.

\end{enumerate}

We have already established that the plane wave (\ref{pw28}) cannot preserve 28 supersymmetries. It remains to find the number
of Killing spinors of the solution when $\mu\not=0$. For this observe that in all $\mu\not=0$ cases, the conditions
({\ref{chop1}}) on $\zeta$  must hold.
On substituting these into ({\ref{bt2}}),  we find
\begin{equation}
\label{cv1}
\bigg( -{1 \over 18} k^2 - {1 \over 9}|\mu|^2-{1\over2}\lambda_1 +{1 \over 24}k \Gamma_{16\sharp} \big(\mu\,
\Gamma_{\bar2\bar3\bar4}+ \bar{\mu}\, \Gamma_{234} \big) \bigg) \zeta =0~.
\end{equation}
Using ({\ref{chop1}}),  we evaluate $V_\alpha\zeta$ to find
\begin{equation}
V_\alpha\zeta=\Gamma_\alpha \bigg( -{1 \over 72}k^2-{1 \over 36} |\mu|^2-{1\over2}\lambda_2
-{1 \over 48}k \Gamma_{16\sharp} \big(\mu
\Gamma_{\bar2\bar3\bar4}+ \bar{\mu} \Gamma_{234} \big) \bigg) \zeta =0
\end{equation}
and $V_{\bar\alpha}\zeta=\bar{(V_\alpha)}\zeta=0$. This calculation is most easily done by taking one value for
$\alpha$ in $V_\alpha\zeta=0$ and repeatedly using ({\ref{chop1}}).  Hence considering both $V_\alpha\zeta=V_{\bar\alpha}\zeta=0$, we find that
\begin{equation}
\label{cv2}
\bigg( -{1 \over 72}k^2-{1 \over 36} |\mu|^2-{1\over2}\lambda_2
-{1 \over 48}k \Gamma_{16\sharp} \big(\mu
\Gamma_{\bar2\bar3\bar4}+ \bar{\mu} \Gamma_{234} \big) \bigg) \zeta =0~.
\end{equation}

It now remains to solve ({\ref{cv1}}) and ({\ref{cv2}}). Indeed, if the kernel of $V_a$ is not trivial,
 ({\ref{cv1}}) and ({\ref{cv2}}) are equivalent to
\begin{equation}
\label{num1}
{1 \over 12}k^2+{1 \over 6}|\mu|^2+{1\over2}\lambda_1 +\lambda_2 =0
\end{equation}
and
\begin{equation}
\label{num2}
\bigg( 2 \lambda_2-{1\over2}\lambda_1 +{1 \over 8} k \Gamma_{16\sharp} \big(\mu
\Gamma_{\bar2\bar3\bar4}+ \bar{\mu} \Gamma_{234} \big) \bigg) \zeta =0 \ .
\end{equation}
This expression can be simplified further using the identity
\begin{equation}
\Gamma_{16\sharp} \phi = i \Gamma_{2\bar2} \phi
\end{equation}
which also follows from ({\ref{chop1}}),
to give
\begin{equation}
\label{num3}
\bigg( 2 \lambda_2-{1\over2}\lambda_1 +{i \over 8} k \big(-\mu
\Gamma_{\bar2\bar3\bar4}+ \bar{\mu} \Gamma_{234} \big) \bigg) \phi =0 \ .
\end{equation}
On squaring ({\ref{num3}}) one finds that
\begin{equation}
({1\over2}\lambda_1-2\lambda_2)^2 \zeta = -{1 \over 8}k^2 |\mu|^2 \zeta \ .
\end{equation}
So if $\zeta\not=0$, then it follows that either  $k=0$ or $\mu=0$. Since we have assumed that $\mu\not=0$, we   shall take $k=0$.
In such a case $V_a$ has a non-trivial kernel provided that
\begin{equation}
k=0, \qquad \lambda_1 = -{2 \over 9}|\mu|^2, \qquad \lambda_2 = -{1 \over 18} |\mu|^2
\end{equation}
and with $\zeta$ satisfying (\ref{chop1}). These conditions are equivalent to the projections
\begin{equation}
\Gamma_{2378} \eta_- =\eta_-~,~~~\Gamma_{2479}\eta_-=\eta_-~.
\end{equation}
 Since the above projections commute with  the equation for $\eta_-$ in (\ref{st2}),  such plane wave solutions preserve 20 supersymmetries.
 These solutions have  been found before in \cite{hg}.
One can easily show that  the Einstein equations and the gauge field equations of eleven dimensional
 supergravity are also satisfied.
To summarize, the plane wave solution of (\ref{pw28}) preserves either 16, or 20, or 32 supersymmetries
depending on the choice of parameters.

\end{document}